%% file: paper.tex
\documentclass[sigconf, authorversion]{acmart}

\usepackage{cleveref}
\usepackage{xspace}
\usepackage{enumitem}

\usepackage{tabularray}
\usepackage{framed}
\usepackage{fancybox,array}
\usepackage{acmart-taps,stfloats}
\def\toprule{\hline}
\def\midrule{\hline}
\def\bottomrule{\hline}
\colorlet{shadecolor}{lightgray}
\colorlet{framecolor}{black}

{\endMakeFramed}

\aptLtoX[graphic=no,type=html]{\newenvironment{froval}{%
\begin{framed}}%
{\end{framed}}}{\newenvironment{froval}{%
\MakeFramed {\advance\hsize-\width\FrameRestore}}%
{\endMakeFramed}}

\crefformat{section}{\S#2\color{blue}#1#3} % see manual of cleveref, section 
\crefformat{subsection}{\S#2\color{blue}#1#3}
\crefformat{subsubsection}{\S#2#1#3}

\newcounter{sfinding}[section]

\newcounter{cfinding}[section]
\newenvironment{cfinding}[1][]{\refstepcounter{cfinding}\par\medskip
   \noindent\textbf{Finding~\thecfinding. #1} \rmfamily}{\medskip}

\AtBeginEnvironment{quote}{\itshape}

\copyrightyear{2024}
\acmYear{2024}
\setcopyright{rightsretained}
\acmConference[CHI '24]{Proceedings of the CHI Conference on Human Factors in Computing Systems}{May 11--16, 2024}{Honolulu, HI, USA}
\acmBooktitle{Proceedings of the CHI Conference on Human Factors in Computing Systems (CHI '24), May 11--16, 2024, Honolulu, HI, USA}
\acmDOI{10.1145/3613904.3642333}
\acmISBN{979-8-4007-0330-0/24/05}

\begin{document}

\title[An In-Depth Study of Online Platforms' Content Moderation Policies]{``Community Guidelines Make this the Best Party on the Internet'': 
An In-Depth Study of Online Platforms' Content Moderation Policies}

\author{Brennan Schaffner}\authornote{The first two authors contributed equally to this research.}

\author{Arjun Nitin Bhagoji}\authornotemark[1]

\author{Siyuan Cheng}

\author{Jacqueline Mei}

\author{Jay L. Shen}

\affiliation{%
  \institution{University of Chicago}
  \city{Chicago}
  \state{Illinois}
  \country{USA}
}

\author{Grace Wang}

\author{Marshini Chetty}

\author{Nick Feamster}

\author{Genevieve Lakier}

\author{Chenhao Tan}
\affiliation{%
  \institution{University of Chicago}
  \city{Chicago}
  \state{Illinois}
  \country{USA}
}

\renewcommand{\shortauthors}{Schaffner and Bhagoji, et al.}

\input{sections/abstract}

%%
%% The code below is generated by the tool at http://dl.acm.org/ccs.cfm.
%% Please copy and paste the code instead of the example below.
%%


%%
%% Keywords. The author(s) should pick words that accurately describe
%% the work being presented. Separate the keywords with commas.

\maketitle
%\renewcommand*{\thefootnote}{\small\ensuremath{\ast}}
%\footnotetext{The first two authors contributed equally to this research.}
%\renewcommand*{\thefootnote}{\arabic{footnote}}

\section{Introduction}\label{sec:intro}
\input{sections/intro.tex}

\section{Background}\label{sec:back}
\input{sections/background.tex}

\section{Research Scope}\label{sec:frame}
\input{sections/framing.tex}

\section{Creating and Annotating the Dataset}\label{sec:method}
\input{sections/method.tex}

\section{Dataset} \label{sec:dataset}
\input{sections/dataset.tex}

\section{Findings}\label{sec:analysis}
\input{sections/analysis.tex}

\section{Discussion and Future Directions}\label{sec:discussion}
\input{sections/discussion.tex}

\input{sections/conclusion.tex}

\begin{acks}
\input{sections/acks.tex}
\end{acks}

\bibliographystyle{ACM-Reference-Format}
\bibliography{chi23}

% \newpage

% \appendix

% \input{sections/appendix.tex}

\end{document}

\appendix

These Supplementary Materials accompany the paper:

\textit{Brennan Schaffner, Arjun Nitin Bhagoji, Siyuan Cheng, Jacqueline Mei, Jay
L. Shen, Grace Wang, Marshini Chetty, Nick Feamster, Genevieve Lakier,
and Chenhao Tan.}
\textbf{“Community Guidelines Make this the Best Party on
the Internet”: An In-Depth Study of Online Platforms’ Content Moderation
Policies.}
In Proceedings of the CHI Conference on Human Factors in Computing
Systems (CHI ’24), May 11–16, 2024, Honolulu, HI, USA. \url{https://doi.org/10.1145/3613904.3642333}

In this document, we provide additional details for our webscraper implementation and tables with more detailed breakdowns of our findings. 

The Supplementary Materials is organized as follows:
\begin{enumerate}
	\item Illustration of the logical flow of our webscraper. (\S~\ref{scraper_logic})
	\item Table containing the manually curated links we used to seed our scraper, by moderation topic and by area (\S~\ref{seed_links})
	\item Table containing the allow/blocklist of patterns used for determining whether to scrape a webpage, by platform.~(\S~\ref{allowblock_list})
	\item Descriptive metrics (number of coded segments, etc.) for the dataset broken down by platform.~(\S~\ref{tab:platformocmpmetrics})
	\item The number of occurrences for each code by platform.~(\S~\ref{sec:cf_by_platform})
	\item The number of occurrences for each code by moderation topic.~(\S~\ref{sec:cf_by_topic})
	\item The equations used to derive the statistical findings for Section 6 of the main paper.~(\S~\ref{sec:percent_equations})
 	\item Alternative analysis for statistical findings that weight platforms equally.~(\S~\ref{alternative_equations})
\end{enumerate}

%71.  Further background and related work (Section A)82.  Exploring benign representations from robust networks (Section B)93.  Analysis of adversarially perturbed representations (Section C)104.  Impact of training parameters on layer-wise convergence (Section D)115.  Alignment of different threat models (Section E)126.  Additional experiments for rebuttal period (Section F)13Our  anonymized  code  is  available  athttps://anonymous.4open.science/r/RobustRS_14NeurIPS_submission-CCF8/README.md.
\newpage
\section{Further Scraper Details}

\subsection{Scraper Logic}
\label{scraper_logic}

\begin{figure}[H]
    \includegraphics[width=0.8\columnwidth]{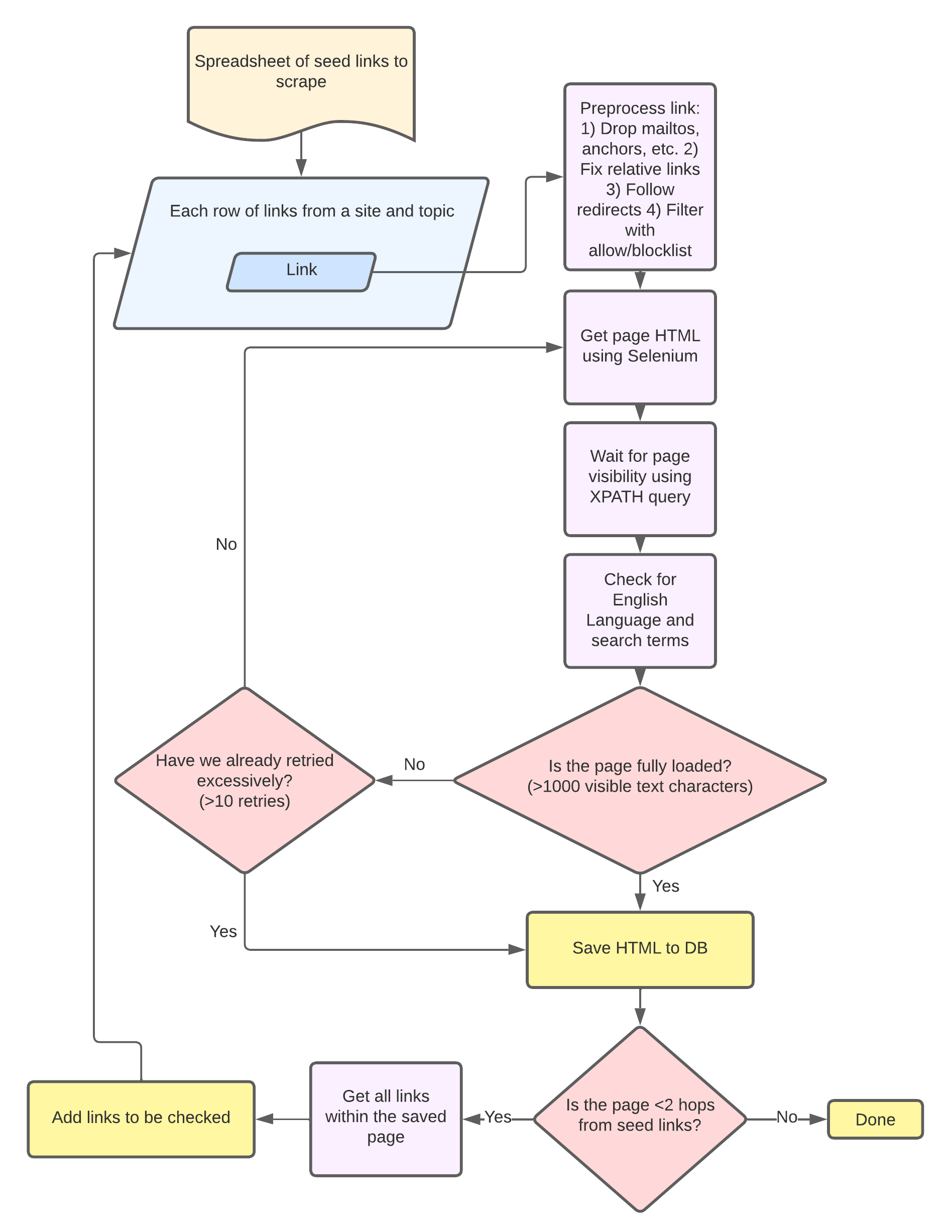}
    \caption{Logical Flow of the Web Scraper.}
    \label{fig:scraperlogic}
\end{figure}

\newpage
\subsection{Seed Links}
\label{seed_links}

\begin{longtable}{l|l}
\caption{The links we used to seed the web scraper, by moderation topic and platform.\label{tab:seedlinks}}\\
\multicolumn{2}{l}{COPYRIGHT INFRINGEMENT}                                                                                                                                                                                                                                                                                                                                                                                                                                                                                                                                                                                                                                                                                                                                                                                                                                                                                                                                                                                                                                                                                                                                                         \endfirsthead 
\hline
\vcell{\textbf{Platform}}                                & \vcell{\textbf{Seed Links}}                                                                                                                                                                                                                                                                                                                                                                                                                                                                                                                                                                                                                                                                                                                                                                                                                                                                                                                                                                                                                                                                                                             \\
\end{longtable}

{
\setlength{\extrarowheight}{0pt}
\addtolength{\extrarowheight}{\aboverulesep}
\addtolength{\extrarowheight}{\belowrulesep}
\setlength{\aboverulesep}{0pt}
\setlength{\belowrulesep}{0pt}
\refstepcounter{table}
\label{tab:seedlinks_misinfo}
\begin{longtable}{l|l} 
%\toprule
\multicolumn{2}{l}{MISLEADING CONTENT}                                                                                                                                                                                                                                                                                                                                                                                                                                                                                                                                                                                                                                                                                                                                                                                                                                                                                                                                                                                                                                                                                                                                                                                                                                                                                                                                                                                                                                                                                                                                                                                                                                                                                                                                                                                                                                                                                                                                                                                    \endfirsthead 
\hline
\vcell{\textbf{Platform}}                               & \vcell{\textbf{Seed Links}}                                                                                                                                                                                                                                                                                                                                                                                                                                                                                                                                                                                                                                                                                                                                                                                                                                                                                                                                                                                                                                                                                                                                                                                                                                                                                                                                                                                                                                                                                                                                                                                                                                                                                                                                                                                                                                                                                                                                     \\
\bottomrule
\end{longtable}
}

%%%%%%%%%%%%%%%%%%%%%%%%%%%%%%%%%%%%%%%%%%

% \usepackage{longtable}
% \usepackage{vcell}
% \usepackage{colortbl}

\refstepcounter{table}
\label{tab:seedlinks_hate}
\begin{longtable}{l|l}
\multicolumn{2}{l}{HARMFUL SPEECH}                                                                                                                                                                                                                                                                                                                                                                                                                                                                                                                                                                                                                                                                                                                                                                                                                                                                                                                                                                                                                                                                                                                                                                                                                                                                                                                                                                                                                                                                                                                                                                                                                                                        \endfirsthead 
\hline
\vcell{\textbf{Platform}}                               & \vcell{\textbf{Seed Links}}                                                                                                                                                                                                                                                                                                                                                                                                                                                                                                                                                                                                                                                                                                                                                                                                                                                                                                                                                                                                                                                                                                                                                                                                                                                                                                                                                                                                                                                                                                                                                                                                     \\
\hline
\end{longtable}

\newpage
\subsection{Allow/Blocklist}
\label{allowblock_list}

\definecolor{Mercury}{rgb}{0.894,0.894,0.894}
\begin{longtblr}[
  caption = {When scraping, a web page was downloaded only if its URL contained one of the Allowlist strings and none of the Blocklist strings. Note that even sites that passed the allow/blocklist were not necessarily relevant to content moderation of our topics of interest and, thus, were not necessarily coded.},
  label = {tab:allowblocklist},
]{
  cells = {t},
  row{even} = {Mercury},
  vline{2-3} = {-}{},
  hline{1,45} = {-}{0.08em},
}
\textbf{Platform}  & \textbf{Allowlist}                                                                                                                                                                                                                                                 & \textbf{Blocklist}                                                                                                                                                                                                                                                                                                                    \\
facebook.com       & facebook.com, fb.com, meta                                                                                                                                                                                                                                         & {.com/news,	.com/formedia, \\.com/journalismproject}                                                                                                                                                                                                                                                                                  \\
youtube.com        & {support.google.com,~	transparencyreport.google.com, \\policies.google.com, youtube,}                                                                                                                                                                              & {.com/@, /channel, /user/, /feed/, \\/jobs/, /trends/, /shorts/, /topic/}                                                                                                                                                                                                                                                             \\
instagram.com      & {help.instagram.com,	facebook.com/terms,\\	facebook.com/business/,	fb.com, \\facebook.com/formedia}                                                                                                                                                                & ~-                                                                                                                                                                                                                                                                                                                                    \\
twitter.com        & twitter.com                                                                                                                                                                                                                                                        & brand.twitter.com                                                                                                                                                                                                                                                                                                                     \\
linkedin.com       & linkedin.com                                                                                                                                                                                                                                                       & .com/checkpoint, .com/uas                                                                                                                                                                                                                                                                                                             \\
wikipedia.org      & policy, meta.wikimedia, foundation.wikimedia                                                                                                                                                                                                                       & {/Minutes, /Resolution, \\:Contributions, /User, CentralAuth/, \\/Meetings, Wikimedia\_Foundation,\\ /Agendas/, /Archive, /Requests\_for}                                                                                                                                                                                             \\
amazon.com         & {amazon.com/gp/help,aws.amazon.com/aup,\\aws.amazon.com/terms,\\amazonregistry.com/legaldocs}                                                                                                                                                                      & {audible.com,comixology.com,\\bookdepository.com,\\amazon.jobs, woot.com,\\.com/gp/customer-reviews,\\.com/product,\\americanexpress.com,\\goodreads.com,amazon.science,\\brandservices.amazon.sg,\\fabric.com}                                                                                                                       \\
pinterest.com      & pinterest.com                                                                                                                                                                                                                                                      & ~-                                                                                                                                                                                                                                                                                                                                    \\
github.com         & {docs.github.com,bounty.github.com,\\github.com/github/dmca/raw,support.github.com,\\help.github.com,github.com/github/docs}                                                                                                                                       & ~-                                                                                                                                                                                                                                                                                                                                    \\
reddit.com         & {redditinc.com,reddit.zendesk,reddit.com/wiki,\\reddithelp.com,/policies/,www.reddit.com/r/reddit,\\www.reddit.com/r/announcements,\\www.reddit.com/r/modnews,\\www.reddit.com/r/reddit,reddit.com/r/blog/,\\reddit.com/r/automoderator}                           & /login,/user/,/rest                                                                                                                                                                                                                                                                                                                   \\
vimeo.com          & {help.vimeo.com,vimeo.zendesk.com/hc,\\vimeo.com/help,vimeo.com/terms,vimeo.com/privacy,\\vimeo.com/dmca}                                                                                                                                                          & {onetrust.com,onetrustpro.com,\\/blog/post,apple.com,/stock,\\/channels,/live,/ott,/create,\\/ondemand,/careers,/solutions,\\/watch,vimeostatus.com,\\facebook.com,cookiepedia,\\developer.vimeo,/business}                                                                                                                           \\
wordpress.com      & {wordpress.com/tos,wordpress.com/support,\\automattic,com/abuse,terms-of-service}                                                                                                                                                                                  & ~-                                                                                                                                                                                                                                                                                                                                    \\
msn.com            & {/servicesagreement,/community,/enduserterms,\\/law-enforcement-requests-report, /terms-of-use,\\/family-hub,/maps/,blogs.microsoft.com, /legal,\\/policies,microsoft.com, msn.com}                                                                                & {/product,/store/,/p/,/d/,/security,\\dynamics.,/solutions,/story,\\powerbi, /account-billing,\\powerapps, optimizely.com,\\xbox, skype}                                                                                                                                                                                              \\
tiktok.com         & {/legal,/copyright-policy,/terms-of-use,\\support.tiktok.com/en/safety,tiktok.com/safety,\\tiktok.com/community-guidelines,\\tiktok.com/transparency}                                                                                                              & {gnu.org,opensource.org,\\facebook.com, mozilla.org,\\spdx.org,apache.org, pta.org,\\connectsafely.org, 988lifeline.org,\\missingkids.org, vimeo.com,\\crisistextline.org,iwf.org, fosi.org,\\newrelic.com,samhsa.gov,\\teamhalo.org,ca.gov,\\consumerfinance.gov,gavi.org,\\oag.ca.org,apple.com,\\vaccinesafetynet.org,nielson.com} \\
xvideos.com        & info.xvideos.com                                                                                                                                                                                                                                                   & ~-                                                                                                                                                                                                                                                                                                                                    \\
tumblr.com         & tumblr.com                                                                                                                                                                                                                                                         & .com/theme,.com/docs,-credits?                                                                                                                                                                                                                                                                                                        \\
pornhub.com        & pornhub.com/information,help.pornhub.com                                                                                                                                                                                                                           & ~-                                                                                                                                                                                                                                                                                                                                    \\
nytimes.com        & nytimes.com,nytco.com                                                                                                                                                                                                                                              & {/wirecutter,cooking.,/puzzles,\\/games,nytimes.com/20}                                                                                                                                                                                                                                                                               \\
flickr.com         & flickr,smugmug                                                                                                                                                                                                                                                     & {/photos,/explore,/events,\\/commons,/map,apple.com,\\facebook.com,/cameras,\\/prints,/create,/jobs,/account}                                                                                                                                                                                                                         \\
fandom.com         & {fandom.com/wiki/Copyright,fandom.com/wiki/Help:,\\fandom.com/wiki/COPPA,fandom.com/terms-of-use,\\fandom.com/wiki/Licensing,:Designated\_agent,\\fandom.com/wiki/DMCA,\\fandom.com/community-creation-policy,\\fandom.com/privacy-policy,fandom.com/wiki/LGBTQIA} & {action=edit,/articles,:Index,\\Special:Contributions,/Talk,\\reddit.com}                                                                                                                                                                                                                                                             \\
ebay.com           & {pages.ebay.com/seller-center,ebay.com/help/policies,\\pages.ebay.com/securitycenter,ebayinc.com/impact}                                                                                                                                                           & ~-                                                                                                                                                                                                                                                                                                                                    \\
imdb.com           & imdb.com                                                                                                                                                                                                                                                           & {.com/event,.com/chart,\\developer.imdb}                                                                                                                                                                                                                                                                                              \\
medium.com         & {policy.medium.com,help.medium.com,\\medium.com/policy,/Medium/Policy,@Medium,\\-terms-of-service}                                                                                                                                                                 & /tag                                                                                                                                                                                                                                                                                                                                  \\
soundcloud.com     & soundcloud.com                                                                                                                                                                                                                                                     & ~-                                                                                                                                                                                                                                                                                                                                    \\
aliexpress.com     & {terms.alicdn,customerservice.aliexpress,\\service.aliexpress,campaign.aliexpress,\\rule.alibaba}                                                                                                                                                                  & ~-                                                                                                                                                                                                                                                                                                                                    \\
twitch.tv          & twitch.tv                                                                                                                                                                                                                                                          & blog.twitch.tv,/jobs                                                                                                                                                                                                                                                                                                                  \\
stackoverflow.com  & {stackoverflow.com,stackexchange.com,\\stackoverflow.blog}                                                                                                                                                                                                         & {/login?,.stackoverflow,\\.stackexchange, /questions/,\\/q/,/signup?,/author/}                                                                                                                                                                                                                                                        \\
archive.org        & archive.org/about,help.archive                                                                                                                                                                                                                                     & /details/,blog.archive                                                                                                                                                                                                                                                                                                                \\
theguardian.com    & {theguardian.com/info,theguardian.com/community,\\theguardian.com/help,manage.theguardian.com,\\theguardian.com/gmg,theguardian.com/about,\\theguardian.com/gnm}                                                                                                   & syndication.theguardian                                                                                                                                                                                                                                                                                                               \\
bbc.co.uk          & bbc.co,bbcstudios.com                                                                                                                                                                                                                                              & /news/,/sport/,search.bbc                                                                                                                                                                                                                                                                                                             \\
xhamster.com       & xhamster.com/info,suggestions.xhamster.com/tos                                                                                                                                                                                                                     & /forum                                                                                                                                                                                                                                                                                                                                \\
quora.com          & help.quora.com,quora.com/about                                                                                                                                                                                                                                     & ~-                                                                                                                                                                                                                                                                                                                                    \\
w3.org             & w3.org/Consortium,w3c.github                                                                                                                                                                                                                                       & ~-                                                                                                                                                                                                                                                                                                                                    \\
sourceforge.net    & {slashdotmedia.com/terms-of-use,/documentation,\\/security,/support}                                                                                                                                                                                               & ~-                                                                                                                                                                                                                                                                                                                                    \\
indeed.com         & {indeed.com/legal,indeed.com/support,\\/legal/anti-slavery/,support.indeed.com}                                                                                                                                                                                    & login,my.indeed                                                                                                                                                                                                                                                                                                                       \\
etsy.com           & etsy.com/legal,/seller-handbook                                                                                                                                                                                                                                    & {/c/,/search/,/listing/,/featured/,\\/market/}                                                                                                                                                                                                                                                                                        \\
sciencedirect.com  & {elsevier.com/legal,elsevier.com/about,\\elsevier.com/connect,elsevier.com/authors}                                                                                                                                                                                & .docx                                                                                                                                                                                                                                                                                                                                 \\
booking.com        & {booking.com/content,reviews\_guidelines,\\trust\_and\_safety,termsandconditions,/trust}                                                                                                                                                                           & ~-                                                                                                                                                                                                                                                                                                                                    \\
imgur.com          & {imgur.com/tos,help.imgur,imgur.com/rules,\\https://imgur.com/removalrequest}                                                                                                                                                                                      & ~-                                                                                                                                                                                                                                                                                                                                    \\
spankbang.com      & spankbang.com/info                                                                                                                                                                                                                                                 & ~-                                                                                                                                                                                                                                                                                                                                    \\
researchgate.net   & *                                                                                                                                                                                                                                                                  & ~-                                                                                                                                                                                                                                                                                                                                    \\
washingtonpost.com & {washingtonpost.com/terms, \\washingtonpost.com/info, \\washingtonpost.com/policies, \\washingtonpost.com/discussions }                                                                                                                                            & ~-                                                                                                                                                                                                                                                                                                                                    \\
xnxx.com           & info.xnxx                                                                                                                                                                                                                                                          & ~-                                                                                                                                                                                                                                                                                                                                    
\end{longtblr}

\newpage
\section{Statistics}
\subsection{Platform-wise Metrics for OCMP-43}
\label{tab:platformocmpmetrics}

\begin{longtable}{l|l|l|l}
\caption{Descriptive Metrics for the Dataset OCMP-43, broken down by platform.\label{tab:ocmp_by_platform}}\\ 
\toprule
\textbf{Platform}                   & \multicolumn{1}{c|}{\textbf{No. Coded Segments}} & \multicolumn{1}{c|}{\textbf{Coded/Total Pages}} & \multicolumn{1}{c}{\textbf{Coded/Total Characters}}  \endfirsthead 
\hline
facebook                            & 599                                              & 73/513                                          & 75,216/1,802,050                                     \\ 
\hline
youtube                             & 1062                                             & 127/437                                         & 132,524/2,234,003                                    \\ 
\hline
instagram                           & 516                                              & 46/159                                          & 56,456/592,670                                       \\ 
\hline
twitter                             & 733                                              & 63/704                                          & 116,440/2,832,162                                    \\ 
\hline
linkedin                            & 499                                              & 77/1174                                         & 66,680/1,906,038                                     \\ 
\hline
wikipedia                           & 179                                              & 26/116                                          & 28,427/548,884                                       \\ 
\hline
amazon                              & 136                                              & 14/21                                           & 15,308/151,278                                       \\ 
\hline
pinterest                           & 141                                              & 16/26                                           & 15,843/171,086                                       \\
\bottomrule
\end{longtable}

\newpage
\begin{landscape}
\subsection{Platform-Wise Breakdown of Code Occurrences}
\label{sec:cf_by_platform}

\begin{longtable}{l|llllll}
\caption{Number of Coded Segments by Platform}\\ 
\hline
\textbf{~} & \multicolumn{1}{c}{\textbf{facebook}} & \multicolumn{1}{c}{\textbf{youtube}} & \multicolumn{1}{c}{\textbf{instagram}} & \multicolumn{1}{c}{\textbf{twitter}} & \multicolumn{1}{c}{\textbf{linkedin}} & \multicolumn{1}{c}{\textbf{wikipedia}} \endfirsthead 
\hline
\rowcolor[rgb]{0.894,0.894,0.894} MODERATION CRITERIA \textgreater{}  Definition & 9 & 4 & 1 & 6 & 2 & 3 \\
MODERATION CRITERIA  \textgreater{} Example & 181 & 199 & 124 & 154 & 113 & 43 \\
\rowcolor[rgb]{0.894,0.894,0.894} \begin{tabular}[c]{@{}>{\cellcolor[rgb]{0.894,0.894,0.894}}l@{}}MODERATION CRITERIA  \textgreater{} Exceptions\end{tabular} & 33 & 73 & 26 & 60 & 9 & 3 \\
JUSTIFICATION  \textgreater{}  Community, Trust,  Safety & 41 & 32 & 28 & 44 & 51 & 22 \\
\rowcolor[rgb]{0.894,0.894,0.894} JUSTIFICATION  \textgreater{}  Legal & 26 & 36 & 25 & 13 & 26 & 3 \\
SAFEGUARDS   \textgreater{} Active User Role & 40 & 195 & 88 & 76 & 68 & 24 \\
\rowcolor[rgb]{0.894,0.894,0.894} SAFEGUARDS \textgreater{}  Platform Detection Methods / Prevention Initiatives & 44 & 130 & 37 & 22 & 44 & 10 \\
BINDING LEGALESE  \textgreater{}  Liability & 3 & 10 & 4 & 7 & 6 & 0 \\
\hline
\end{longtable}

\newpage

\begin{longtable}{lllllllllllllll}
\caption*{Number of Coded Segments by Platform (cont.)}\\ 
\hline
\multicolumn{1}{c}{\textbf{amazon}}                                & \multicolumn{1}{c}{\textbf{pinterest}} & \multicolumn{1}{c}{\textbf{github}} & \multicolumn{1}{c}{\textbf{reddit}} & \multicolumn{1}{c}{\textbf{vimeo}} & \multicolumn{1}{c}{\textbf{wordpress}} & \multicolumn{1}{c}{\textbf{msn}} & \multicolumn{1}{c}{\textbf{tiktok}} & \multicolumn{1}{c}{\textbf{xvideos}} & \multicolumn{1}{c}{\textbf{tumblr}} & \multicolumn{1}{c}{\textbf{pornhub}} & \multicolumn{1}{c}{\textbf{nytimes}} & \multicolumn{1}{c}{\textbf{flickr}} & \multicolumn{1}{c}{\textbf{fandom}} & \multicolumn{1}{c}{\textbf{ebay}}  \endfirsthead 
\hline
\rowcolor[rgb]{0.894,0.894,0.894} 0                                & 1                                      & 1                                   & 9                                   & 5                                  & 1                                      & 4                                & 13                                  & 0                                    & 0                                   & 2                                    & 0                                    & 0                                   & 0                                   & 4                                  \\
39                                                                 & 96                                     & 95                                  & 60                                  & 78                                 & 49                                     & 21                               & 106                                 & 8                                    & 91                                  & 69                                   & 35                                   & 28                                  & 29                                  & 69                                 \\
\rowcolor[rgb]{0.894,0.894,0.894} 8                                & 2                                      & 15                                  & 14                                  & 11                                 & 20                                     & 13                               & 22                                  & 0                                    & 6                                   & 4                                    & 1                                    & 1                                   & 7                                   & 19                                 \\
\hline
\end{longtable}

\newpage

\begin{longtable}{lllllllllllll}
\caption*{Number of Coded Segments by Platform (cont.)}\\ 
\hline
\multicolumn{1}{c}{\textbf{imdb}}                                 & \multicolumn{1}{c}{\textbf{medium}} & \multicolumn{1}{c}{\textbf{soundcloud}} & \multicolumn{1}{c}{\textbf{aliexpress}} & \multicolumn{1}{c}{\textbf{twitch}} & \multicolumn{1}{c}{\textbf{stackoverflow}} & \multicolumn{1}{c}{\textbf{archive}} & \multicolumn{1}{c}{\textbf{theguardian}} & \multicolumn{1}{c}{\textbf{bbc}} & \multicolumn{1}{c}{\textbf{xhamster}} & \multicolumn{1}{c}{\textbf{quora}} & \multicolumn{1}{c}{\textbf{w3}} & \multicolumn{1}{c}{\textbf{sourceforge}}  \endfirsthead 
\hline
\rowcolor[rgb]{0.894,0.894,0.894} 0                               & 0                                   & 13                                      & 1                                       & 2                                   & 1                                          & 0                                    & 0                                        & 8                                & 0                                     & 2                                  & 14                              & 0                                         \\
12                                                                & 70                                  & 70                                      & 48                                      & 102                                 & 24                                         & 5                                    & 23                                       & 51                               & 19                                    & 25                                 & 31                              & 5                                         \\
\rowcolor[rgb]{0.894,0.894,0.894} 0                               & 5                                   & 9                                       & 1                                       & 12                                  & 1                                          & 2                                    & 3                                        & 20                               & 0                                     & 2                                  & 1                               & 0                                         \\
\hline
\end{longtable}

\newpage

\begin{longtable}{lllllllll}
\caption*{Number of Coded Segments by Platform (cont.)}\\ 
\hline
\multicolumn{1}{c}{\textbf{indeed}}                                & \multicolumn{1}{c}{\textbf{etsy}} & \multicolumn{1}{c}{\textbf{sciencedirect}} & \multicolumn{1}{c}{\textbf{booking}} & \multicolumn{1}{c}{\textbf{imgur}} & \multicolumn{1}{c}{\textbf{spankbang}} & \multicolumn{1}{c}{\textbf{researchgate}} & \multicolumn{1}{c}{\textbf{washingtonpost}} & \multicolumn{1}{c}{\textbf{xnxx}}  \endfirsthead 
\hline
\rowcolor[rgb]{0.894,0.894,0.894} 8                                & 0                                 & 1                                          & 0                                    & 0                                  & 0                                      & 0                                         & 0                                           & 0                                  \\
73                                                                 & 112                               & 26                                         & 32                                   & 40                                 & 15                                     & 31                                        & 17                                          & 9                                  \\
\rowcolor[rgb]{0.894,0.894,0.894} 0                                & 20                                & 7                                          & 0                                    & 9                                  & 1                                      & 1                                         & 0                                           & 0                                  \\
15                                                                 & 29                                & 16                                         & 13                                   & 22                                 & 2                                      & 12                                        & 6                                           & 3                                  \\
\rowcolor[rgb]{0.894,0.894,0.894} 24                               & 35                                & 8                                          & 0                                    & 7                                  & 5                                      & 14                                        & 0                                           & 4                                  \\
\hline
\end{longtable}

\end{landscape}
\newpage
\subsection{Topic-Wise Breakdown of Code Occurrences}
\label{sec:cf_by_topic}

\begin{longtable}{l|l|l|l}
\caption{Number of Coded Segments by Moderation Topic}\\ 
\toprule
~                                                                                                                          & \multicolumn{1}{c|}{\begin{tabular}[c]{@{}c@{}}\textbf{COPYRIGHT }\\\textbf{INFRINGEMENT}\end{tabular}} & \multicolumn{1}{c|}{\begin{tabular}[c]{@{}c@{}}\textbf{HARMFUL}\\\textbf{SPEECH}\end{tabular}} & \multicolumn{1}{c}{\begin{tabular}[c]{@{}c@{}}\textbf{MISLEADING}\\\textbf{CONTENT}\end{tabular}}  \endfirsthead 
\hline
MODERATION CRITERIA \textgreater{} Definition                                                                                            & 31                                                                                                      & 51                                                                                                 & 33                                                                                             \\ 
\hline
MODERATION CRITERIA  \textgreater{}

Example                                                                                             & 384                                                                                                     & 725                                                                                                & 1410                                                                                           \\ 
\hline
MODERATION CRITERIA  \textgreater{} 

Exceptions                                                                                          & 187                                                                                                     & 124                                                                                                & 130                                                                                            \\ 
\hline
JUSTIFICATION \textgreater{} Community, Trust,  Safety                                                                                   & 114                                                                                                     & 390                                                                                                & 333                                                                                            \\
\bottomrule
\end{longtable}

\newpage

\input{appendix_equations}

%% file: sections/abstract.tex
\begin{abstract}
%Content moderation on online platforms that host content posted by users
%(``user-generated content'') is critical for balancing user safety with
%freedom of speech. 
Moderating user-generated content on online platforms is crucial for balancing
user safety and freedom of speech. Particularly in the United States, platforms
are not subject to legal constraints prescribing permissible content. Each
platform has thus developed bespoke content moderation policies, but there is
little work towards a comparative understanding of these policies across
platforms and topics. This paper presents the first systematic study of these
policies from the 43 largest online platforms hosting user-generated content,
focusing on policies around copyright infringement, harmful speech, and
misleading content. We build a custom web-scraper to obtain policy text and
develop a unified annotation scheme to analyze the text for the presence of
critical components. We find significant structural and compositional variation
in policies across topics and platforms, with some variation attributable
to disparate legal groundings. We lay the groundwork for future studies of
ever-evolving content moderation policies and their impact on users.
\end{abstract}

%% file: sections/intro.tex
As much of the world's discourse happens online, platforms that host and
mediate online content play an increasingly influential role in moderating
societal conversations. Platforms like Twitter, Threads, and TikTok have been
compared to ``global town squares''~\cite{tiktok_replace_twitter}. The
practice of deciding whether to publish, remove, and flag content that is
posted by third-party users is typically referred to as {\em content
moderation}~\cite{kiesler2011regulating,keller2020facts}. 
This is what Goldman~\cite{goldman2021content} refers to as ``content regulation''. 
% Due to the critical importance of platforms' policies in influencing online 
% behavior for millions of users, as well as shaping societal discourse, 
% it has become critically important to understand how content moderation
% policies are structured and what they contain.
% Because platform policies influence online behavior for 
% millions of users and shape societal discourse, it is crucial to understand
% how content moderation policies are structured and what they contain. 
Since platform policies influence online behavior for 
millions of users and shape societal discourse, it is crucial to understand
how content moderation policies are structured and what they contain. 

Platforms that are faced with the prospect of moderating content face two
primary challenges: (1)~enforcing policies at scale; (2)~ensuring that policies
are applied consistently. First, as the {\em scale} of data to be monitored and
moderated has increased, consistent content moderation has become
extremely challenging. In part due to the scale of the problem, alongside regulatory
pressures, platforms have increasingly attempted to rely on algorithmic
automated content moderation~\cite{gorwa2020algorithmic}. Yet, in reality,
automation has not been able to replace human decision-making: much content
moderation is implemented by underpaid contractors, largely in the Global
South~\cite{facebook_content_mod}, hired by large
platforms~\cite{meta_content_mod}. A second goal of content moderation is {\em
consistency} in how the policies are applied---across instances of content,
users, geographies, and so forth. Greater consistency may lead to increased
trust and improved discourse~\cite{kiesler2011regulating}. Unfortunately, public
perception is that content moderation is inconsistently
implemented~\cite{sabin23,ozanne2022shall}. \citet{keller2020facts} also show it
is often difficult to determine where, how and according to what rules content
moderation is occurring.

% \cite{keller2020facts}.
% particularly with regard to polarizing topics. Critics often claim political
% agendas motivate differences in which users and/or content are targeted by
% platform moderation\todo{cite}, although this is a contested
% narrative\todo{cite}. Previous research has also pointed out that it ~
% Smaller platforms without the operational
% budgets to invest in complex and expensive content moderation systems typically
% struggle to keep pace with the increasing rates of posted online content.

% When both the platforms and
% users can both understand and predict when content is likely to be removed or
% flagged, users may ultimately have greater trust in
% the platform, and discourse and exchange of information on the platform may
% ultimately be more trustworthy.

% Differences in content moderation policies can occur both
% by platform (e.g., Facebook vs. Twitter) and by area (e.g., misinformation vs.
% copyright infringement).

In this paper, we seek a unified understanding of what these \emph{rules} are in
the first place. We look at how many online platforms specify their content
moderation policies, with an in-depth study and structured analysis across a
wide range of platforms. Fortunately, many platforms typically provide some
level of transparency into their content moderation policies, allowing us to
study them in some level of detail (including, for example, what is specified
vs. unspecified, and how the structure and content of these policies vary across
platforms). With the exception of copyright enforcement, content moderation
largely lacks a prescriptive regulatory approach (particularly in the United
States), leading to potential divergence of policies and inconsistency of
application, even within a single platform. This state of affairs makes it
paramount to study platforms' content moderation policies as they define public
discourse, and being user-facing, influence the manner in which users interact
and conduct themselves online.  

We study content moderation policies for several topics, given that the nature
of content moderation policies typically differs depending on the type of
content. For example, copyright rules have a well-established legal regime,
especially in the United States after the enactment of the Digital Millennium
Copyright Act (DMCA), which makes it an insightful contrast to other types of
content that might be moderated, such as misinformation and hate speech. To
capture the ends of this spectrum in our work, we focus on three content
moderation topics: (1)~copyright infringement, (2)~hate (or harmful) speech, and
(3)~misinformation (or misleading content). 
Each topic represents different intensities of legal grounding
and has entered the public discourse about online content at different times. 
Moreover, the prominence of the three topics makes it likely that
each studied platform will contain policies pertaining to each topic.

This paper performs the first in-depth collection, annotation, and analysis of
content moderation policies, across the 43 largest online platforms (determined
using Tranco~\cite{LePochat2019}) hosting user-generated content. A better
understanding of how content moderation policies are structured and what they
contain may lead to improved alignment in platform policies, regulation, and
user expectations. Towards this goal, we pose the following research questions: 
\begin{enumerate}[nosep]
    \item Collection: How do we systematically collect all text related to
   content moderation policies across a range of platforms?
   \item Annotation: How can we annotate content moderation policy text to
   reveal key components that are important for users and capture similarities
   and differences across topics and platforms?
   \item Analysis: How consistent are content moderation policies in
   structure and composition across different platforms, and how do they
   relate to existing legal frameworks? 
\end{enumerate} 

%\begin{figure}
 %   \includegraphics[width=\columnwidth]{graphics/example_annotatedtext.png}
  %  \caption{Example of a coded policy passage.}
   % \label{fig:policy_passage}
%\end{figure}

The {\em collection} effort is substantial given the relatively unstructured
nature of these policies, and the lack of standardized approaches to expressing
them. Specifically, these policies are often scattered, imprecise and
non-committal (past work has also demonstrated that at least in specific
cases, enforcement may also be inconsistent~\cite{matias2022software}).
Collecting policies across a wide range of platforms in a systematic manner to
create a standardized, accessible dataset is thus an important first step to
enabling both this paper and future work to reach more general conclusions
concerning how platforms approach content moderation. Because data collection
results in a large volume of text, having a clear {\em annotation} scheme to
provide insights about the data at scale, particularly with regard to
information provided to and actions for users of the platform to take.
Finally, standardized and wide-ranging data collection allows for
cross-platform and cross-topic {\em analysis}, which can provide insights about
the role legal regimes, platform size, and other factors play in policy
structure and composition.
This paper presents the following contributions:

\vspace*{0.1in}
\noindent
\textbf{1. Open-source collection pipeline enabling continual
collection~(\cref{subsec:corpus_creation}).} We create an open-source
pipeline\footnote{The repository can be found here: \href{https://anonymous.4open.science/r/content-moderation-7AC0/}{https://anonymous.4open.science/r/content-moderation-7AC0/README.md}} to collect and build a dataset of
content moderation policies. The pipeline consists of a scraper and text
extractor. Our scraper overcomes common hurdles for web-scrapers such as
bot-blocking, dynamic page loading via javascript and rate limiting. The scraper
includes an iterative process to obtain relevant policy text across a platform
using parsing and keyword search.

% The final portion of the pipeline extracts
% relevant text from within webpages to reduce irrelevant text. 

\vspace*{0.1in}
\noindent
\textbf{2. Inductively-designed policy annotation
scheme~(\cref{subsec:codebook}) :} We develop an annotation scheme (or codebook)
that captures critical components of policies from a user perspective. The codebook, developed in consultation with
legal experts, takes an intent-driven and user-centric approach to the text,
aiming to highlight the purpose and the relevance of policy statements. It
enables further mixed-methods analysis to bring out policy differences across
platforms and topics, in terms of communication clarity and the impact of legal
regimes.

% to qualitatively analyze our dataset of policies. The codebook  The codebook is also designed to
% bring out differences across platforms in terms of their focus on motivating and
% clearly communicating policy to users. It also serves to highlight how the
% existence of legal regimes impacts the actions users can take when interacting
% with content that can be moderated.

\vspace*{0.1in}
\noindent
\textbf{3. Dataset of annotated policies~(\cref{sec:dataset}):} The scraping and
annotation processes result in a dataset we name OCMP-43 (Online Content
Moderation Policies, set of 43 platforms) for convenience. The dataset consists of over
$1000$ annotated pages of policy text with tens of thousands of annotated
segments.\footnote{This paper's title is adapted from Imgur's policy pages.}
We provide the open-source dataset, including annotated text, to
enable further research into existing content moderation policies.\footnote{The dataset is made available at \url{https://ocmp43.cs.uchicago.edu}.} We show that
our dataset contains policy text scattered across diverse areas of a platform's
pages, further underscoring the value of our consolidation.

\vspace*{0.1in}
\noindent
\textbf{4. Mixed-methods analysis of policy content, structure and
coverage~(\cref{sec:analysis}):} We use our annotated dataset to analyze content
moderation policies at scale. Our key questions concern platforms' stated intent
and methods in content moderation policies, as well as possible impact on users.
A few key findings are that (1) platforms rarely explicitly define what they
intend to moderate; (2) platforms, depending on the topic, can rely heavily on
users for moderation; and (3) except in the case of copyright infringement,
users rarely have recourse after being moderated.

The results presented in this paper enable and encourage researchers to perform
further detailed studies of content moderation policies. The custom data
collection pipeline we have developed enables studies of the evolution of
platform policies over time. Our annotated dataset can help highlight best
practices and provide recommendations for future policies, as well as to find
policies that may show platforms acting in bad faith. Finally, our dataset and
preliminary findings lay the groundwork for future user studies to understand
the interaction of users' with these policies, as well as large-scale audits to
determine when policies are being consistently implemented.

%% file: sections/background.tex
In this section, we first provide a brief history of online content moderation,
with a focus on the legal and policy regime in the United States. This helps set
the context for why analyzing the moderation policies of different platforms is
meaningful, as they often serve as proxy regulators in the absence of focused
government regulation. We then provide a brief survey of related academic work
that has studied content moderation by online platforms.

\subsection{History and overview of online content moderation}
Since the development of the world wide web as a medium for information
exchange, there has been a proliferation of online platforms where people gather
to share information and ideas. Their evolution can be traced from the
simplicity of early message boards, where users could linearly post text to
respond to each other, to modern platforms like Twitch and Facebook where live
video can be shared even as users comment on it. As these platforms have grown
in scale and scope, the content being shared on them has begun to have
real-world impact, including, in some cases, the incitement of
violence~\cite{mackintosh21,amnesty22}.

However, governments, particularly in the Global North, have been reluctant to
hold platforms responsible for the content that is posted on them, out of
fear that doing so will lead platforms to either take down too much valuable
content, or refrain from moderating content altogether. In the United States,
Section 230 of the Title 47 of the United States Code\footnote{This
defined the structure and role of the Federal Communications Commission (FCC),
the regulatory body that deals with communication via modern technologies such
as radio and satellite.}, explicitly
provides platforms protection from liability for the third-party speech that
they publish. This Section also protects platforms from liability for the
``good faith'' moderation of third-party content they deem objectionable. The
result has been to make it difficult to impose legal liability on platforms for
much of the speech that appears on their websites. In the absence of clear
rules, detailing what they should do about harmful or controversial content,
platforms have developed elaborate policies to guide their regulation of speech.

These rules are intended to guide users, and demonstrate to members of the
general public that the platforms are exercising their power over speech
responsibly. Despite these aims, it is often very difficult for either users or
researchers to know how platforms are applying these policies, and whether they
are doing so consistently. This is notwithstanding calls from civil society
organizations for many years now for clear and consistent decision-making by the
stewards of the digital public sphere~\cite{santaclara18}. Consistency and
clarity are understood to be important goods in themselves—one of the rights
that users are entitled to, when they speak on the internet—and also as a means
of ensuring the legitimacy and thus the effectiveness of content moderation
decision-making. The theory here is that consistently applied moderation
criteria help regulate and make online communities safe for their participants,
and also “increase the legitimacy and thus the effectiveness of moderation
decisions''~\cite{kiesler2011regulating}.

In this paper, we seek to understand what
comprises moderation criteria for modern online platforms, as espoused by the
platforms themselves. This is important for users, since in the absence of
unified legal frameworks governing their behavior online, each platform's stated
moderation criteria are all the user can use as a guide.

\subsection{Related Work}

\noindent \textbf{Content moderation in online communities:} Content moderation
has attracted substantial interest from computer science researchers because of
its growing importance in online communities~\cite{kiesler2011regulating}.
Content moderation can potentially filter ``bad'' content and ``bad'' actors and
promote a healthy community, but may also infringe an individual's right to
freedom of expression. Thus, there is a large body of research studying the
effect of moderation on community behavior, including whether one should
regulate online content at
all~\citep{Chancellor:2016:TIC:2818048.2819963,chandrasekharan2017you,srinivasan+dnm+lee+tan:19,jhaver2019does,Chang-Recidivism:19,seering_shaping_2017,matias2019preventing,cheng+dnm+leskovec:2014}.
For instance, prior work has developed an observational method that leverages
delayed feedback (\emph{i.e.}, content moderation does not happen
instantaneously) to understand the causal effect of comment removal on users'
future behavior~\cite{srinivasan+dnm+lee+tan:19}.

There is a also growing line of research on
characterizing the rules on online platforms~\cite{fiesler_reddit_nodate,fiesler_what_2017,
fiesler+al:2018,Brian_keegan_2017,chandrasekharan_internets_2018}, all of which tend to
focus on a single platform. 
For instance, some researchers analyzed \textsf{wikipedia.com}'s
publicly available records and editor discourse to uncover patterns in
rules and rule-making communities~\cite{butler2008don, hwang2022rules} and track
rules over 
time~\cite{Brian_keegan_2017, beschastnikh2008wikipedian}.
% For instance, some work~\cite{Brian_keegan_2017} studied the evolution
% of rules on \textsf{wikipedia.com} by tracking revisions on
% rule-related Wikipedia pages. 
In another
study, researchers \cite{fiesler+al:2018} provided a characterization of
different types of rules and show that community rules share common
characteristics across subreddits. In comparison, other
researchers~\cite{chandrasekharan_internets_2018} performed a large-scale
study to understand content moderation through language used in
removed comments on Reddit and identify norms that are universal (macro),
shared across certain groups (meso), and specific to individuals (micro).  

Our paper differs from this body of work in the scale of the cross-platform
analysis we perform, enabled by a custom web-scraper and policy annotation
scheme. By gathering and analyzing platforms' content moderation policies at
scale, as well as releasing a fully annotated dataset, we set the stage for 
future user and audit studies that go
beyond traditionally studied platforms such as \textsf{reddit.com}.
The publication of our dataset takes some inspiration from \citet{wilson2016creation}, 
who released an annotated dataset of privacy policies. However, our methodology and scope
differ substantially. 
% In particular, they manually downloaded privacy policies, while we employed 
% a custom web-scraper to find and extract dispersed content moderation policies.
In particular, while they manually downloaded privacy policies, we diverged 
in our approach by developing a custom web-scraper to automatically retrieve and extract 
dispersed content moderation policies, showcasing a distinct approach to locating 
relevant policy text across the sites, and employing a novel annotation scheme.
The closest related work to ours is~\citet{singhal2022sok}, which qualitatively surveys 14
different platforms' content moderation policies. Our work, in contrast, enables
quantitative analysis due to our large annotated dataset and only 8 of the
studied platforms overlap, with our work studying 35 additional platforms. 

\noindent \textbf{Legal research on content moderation:} The importance of content
moderation has spurred a robust legal
discussion~\cite{gillespie2018custodians,suzor2019lawless,klonick2017new}. Three
questions are outlined in previous work
\cite{goldman2021content,klonick2017new}: (1) what content should be allowed
online? This question is concerned with the First Amendment of the United States
Constitution. As discussed in \cref{sec:intro}, the laws are much clearer for
copyright infringement than for misleading content; (2) who should make the
substantive rules of online content and activities?
Klonick~\cite{klonick2017new} further distinguishes standards (e.g., ``don't
drive too fast'') from rules (e.g., ``a speed limit set at sixty-five miles per
hour'') in the current practice of online platforms; (3) who should determine
if a rule violation has occurred, and who should hear any appeals of those
decisions?  While existing research is mostly qualitative, our goal is to
examine existing platform rules in detail and at scale. 

\noindent \textbf{User understanding of content moderation:} There is a large
and growing body of research around how users understand and react to content
moderation~\cite{fiesler_understanding_2015, fiesler_reality_2016}. For instance, there are numerous studies that examine how
\textsf{reddit.com}'s users react to post-level and community-level
moderation~\cite{jhaver2019does, jhaver_did_2019, jhaver_human-machine_2019,
jhaver_online_2018, matias_going_2016, chandrasekharan_quarantined_nodate,
chandrasekharan_you_2017, chandrasekharan2019crossmod,
Copeland:vldb1989,jhaver_evaluating_2021}. Prior work has shown that while
banning subreddits with hateful speech does decrease toxicity overall, however,
toxicity within the banned subreddit may increase
\cite{chandrasekharan_you_2017, Copeland:vldb1989}. In some cases, content
moderation can increase toxicity and move users to more extreme behavior on
other platforms~\cite{horta_ribeiro_platform_2021}. Some studies also imply that
users who know about community rules or who received explanations for their
removals are often more accepting of moderation practices \cite{jhaver_did_2019,
jhaver2019does}. Some user studies have also studied content moderators
themselves~\cite{roberts_commercial_nodate, roberts2014behind, ahmad_its_2019}.

Researchers have also examined how users react to algorithmic content
moderation~\cite{vaccaro_at_2020} and softer forms of content moderation such as
and shadow banning \cite{myers_west_censored_2018, zannettou_i_2021}. Most of
this work focused on content moderation on a single platform, for a single
topic. In contrast, our work aims to create an understanding of content
moderation policies across multiple platforms by synthesizing what users see
when trying to understand how they may be moderated.

%% file: sections/framing.tex
In this section, we describe the research scope within which we analyze the content
moderation policies of different websites. At a high level, we focus on three
different topics across 43 of the 200 top websites from the
Tranco~\cite{LePochat2019} list\footnote{Tranco provides a
manipulation-resistant and publicly available list of the top websites for
researchers. The list we use is accessible at
\url{https://tranco-list.eu/list/Q9X24/full}.} generated on $29$ June $2022$,
which we determine to host user-generated content.

\subsection{Topics of focus}
Modern platforms host a wide range of content that may have to be regulated,
either due to explicit legal frameworks that require it, or due to the
platforms' own motivations, which could stem from ethical or economic
considerations, or both. In this paper, we focus on three broad topics within
which content that tends to get regulated often falls: \emph{copyright
infringment}, \emph{harmful speech}, and \emph{misinformation and misleading
content}. Our choice of these three topics is motivated by three factors. First,
we find policy pertaining to these three topics is highly prevalent across all the
websites we consider, highlighting their importance and ubiquity. This is in
contrast to niche or newly emerging topics that may not appear in all platform
policies (\emph{e.g.}, regulations concering deepfakes or Generative Artificial
Intelligence). Second, these topics have been the focus of much recent debate
and consideration, both in the public and academic spheres, due to their
potential for direct harm to platform
users~\cite{amnesty22,loomba2021measuring}. Finally, copyright infringement
offers a meaningful contrast to the other two topics, due to the presence of
strong copyright laws in many countries while the other content areas have fewer
governmental regulations. We are aware that there are other areas such as child
pornography where moderation is common, but we limit our scope in this paper for
focus, and to avoid the additional complications of exploring these areas with
complex ethical considerations.

\subsubsection{Copyright Infringement} 
There are complex legal regimes that regulate intellectual property around the
world, with a particularly prominent one being the Digital Millennium Copyright
Act (DMCA)~\cite{congress1998digital} in the United States. Although the DMCA does limit platforms' liability
for copyright violations by content posted on them, they nonetheless tend to
have well-defined policies that specify actions users can take with regard to
copyrighted content.

\subsubsection{Harmful speech}
Naturally, platforms may restrict content that is directly harmful to
other users of the platform, as this can reduce user engagement
\cite{kiesler2011regulating}, leading to direct economic concerns. We refer to
content of this form as \emph{abuse}. In addition to content that directly
targets other users, there is an increasing prevalence
\cite{gorwa2020algorithmic} of content that is offensive towards groups of
people sharing characteristics, which is broadly referred to as \emph{hate
speech}. We group these two types of content under the area of harmful speech.

In many contexts, it is challenging to distinguish harmful speech from
provocative or satirical speech, especially in light of cultural norms that vary
both geographically and temporally. A lack of clear definitions of what
constitutes harmful speech---and its increasing scale in recent years---has made
it challenging to moderate. In the United States in particular, the First
Amendment to the Constitution protects speech in a variety of contexts, leading
to the absence of clear legal principles within which harmful speech can be
regulated. This legal regime is in contrast to Germany, for example, where the
Network Enforcement Act (NetzDG)~\cite{knight18} explicitly requires platforms
to take down hate speech. There is thus significant variation in how platforms
deal with the presence of harmful speech, making it an interesting area to
study. 

\subsubsection{Misinformation and Misleading Content}
As the reach of online platforms increases, the impact of and trust in
conventional sources of information has eroded. This came to light in recent
political campaigns in the United States and India, as well as during the course
of the COVID-19 pandemic. However, the ease of spread of misinformation on
online platforms has had debilitating consequences on public health and safety,
making it a critical area for moderation for
platforms~\cite{loomba2021measuring}. The polarization regarding what is
misinformation or not, particularly in light of mistrust of traditional
institutions like universities and the government, makes it challenging for
platforms to moderate without alienating portions of its user base. With rapidly
changing conditions, it is often unclear what the consensus is on certain
critical topics, which makes it difficult to moderate in real-time.

In addition to misinformation regarding public health and safety, we also
categorize spam, fake products, and false advertising under misleading content,
as these are all types of misleading content that platforms need to regulate in order to
not erode user trust. Depending on the type of content hosted on the platform,
there are subtle variations in how misleading content appears (\textit{e.g.,}
clothing with misinformation printed on it is found on Etsy). Throughout the remainder of the work, we use `misleading content' to refer to all of these forms of misleading content and misinformation.

\begin{table}
    \centering
    \caption{The 43 platforms in our dataset of online content moderation policies (OCMP-43), ordered by their Tranco ranking.}
    \label{tab:platformlist}
    \resizebox{\linewidth}{!}{%
    \begin{tabular}{|c|} 
    \toprule
    \textbf{Platforms in OCMP-43} \\ 
    \midrule
    \multicolumn{1}{|l|}{\begin{tabular}[c]{@{}l@{}}facebook.com, youtube.com, instagram.com, twitter.com, linkedin.com, \\wikipedia.org, amazon.com, pinterest.com, github.com, reddit.com, \\vimeo.com, wordpress.com, msn.com, tiktok.com, xvideos.com,\\tumblr.com, pornhub.com, nytimes.com, flickr.com, fandom.com, \\ebay.com, imdb.com, medium.com, soundcloud.com, aliexpress.com,\\twitch.tv, stackoverflow.com, archive.org, theguardian.com, bbc.co.uk,\\xhamster.com, quora.com, w3.org, sourceforge.net, indeed.com, \\etsy.com, sciencedirect.com, booking.com, imgur.com, spankbang.com, \\researchgate.net, washingtonpost.com, xnxx.com\end{tabular}}  \\
    \bottomrule
    \end{tabular}
    }
\end{table}

\subsection{Platforms of focus}
There is a proliferation of online platforms that host user-generated content.
Our focus in this paper is to provide a method to collect content moderation
policies from a wide set of platforms, as well as to collate a large dataset of
policies. We considered the top $200$ websites from the Tranco
list~\cite{LePochat2019} obtained on the $29^{\text{th}}$ of June, 2022\footnote{The list can be accessed at
\url{https://tranco-list.eu/list/Q9X24/full}}. Of these, we filtered out those
websites that do not host user-generated content, such as \url{google.com},
\url{akamaiedge.net} and \url{baidu.com}, as well as those which may host
user-generated content, but are not primarily in English\footnote{This is a
necessary limitation as English is the only language all authors of the paper
are sufficiently proficient in. Regardless, our method for scraping can be
adapted for platforms in other languages with some modification.} such as
\url{bilibili.com}. We also combined websites that correspond to the same
platform, and thus use the same set of content moderation policies, such as
\url{wikipedia.org} and \url{wikimedia.org}. The complete list of the resulting 43 platforms we
consider is presented in Table~\ref{tab:platformlist}.

To determine if a platform hosts user-generated content, and if it is
primarily in English, we use manual analysis. We visited each website in turn,
and checked if any portion of the platform contained user-generated content,
which would in turn indicate the possible presence of policies regarding content
moderation. For example, while \url{booking.com} largely contains links to stay
and transport options for travelers, we include it in our dataset since it also hosts a comment and feedback section where users can interact with each other, governed by platform policies on moderation.

The resulting list of platforms host a diverse spectrum of user-generated
content. For instance, large social media powerhouses like \url{facebook.com},
\url{instagram.com}, and \url{twitter.com} feature an array of multimedia content, including text,
images, videos, and live streams. The variety of media is accompanied by a
variety of intents, from consuming and debating news to friendly banter. On
platforms such as \url{linkedin.com} and \url{github.com}, the focus shifts towards professional
networking and collaborative coding, leading to distinct user generated
content such as resumes and code documentation. The wealth of \url{wikipedia.com}'s content
comes from world-wide volunteer contributions to articles. Sites such as \url{reddit.com}
and \url{tumblr.com} foster semi-siloed communities that generate content based on
specific themes and, in some circumstances, are expected to self-moderate.
Platforms like \url{pornhub.com} and \url{xvideos.com} cater to explicit adult content, posing
unique challenges for content moderation. Meanwhile, news platforms must
govern both their journalists' content and the community's comments, with
heightened expectations of content accuracy. Further, e-commerce platforms
like \url{amazon.com} and \url{etsy.com} take on additional responsibilities regulating both
vendor content (e.g., product listings and FTC advertising regulations) and
buyer content (e.g., reviews and ratings). Our dataset of content moderation
policies captures the tailored content moderation policies necessitated by
these variations in content types and user dynamics. 

% - While the forgoing hints at the possibilities of analysis based on platform categorization, true categorization is a difficult task. 
% - platforms are often multi-faceted, and can be categorized in many, often overlapping ways.
% - the state of existing categorization tools highlights the difficulty of this task. Alexa has been deprecated, Cloudflare puts platforms into multiple overlapping categories
% - as such, we resist analysis relating to platform categorizations. 
% - Except in one case (analysis section), where we specify policy completion for a few categories and we only use the 3 categories with the least dispute as to categorization.
% - in particular, for that part, we use the following categories: adult content contains (pornhun, xhamster, xnxx, spankbang, and xvideos),
%  news containts (nytimes, the guardian, bbc, wapo), and e-commerce contains (amazon, ebay, etsy, aliexpress)

%\todo{one paragraph describing websites in more detail: what kind of content they mainly host, where the user-generated content on that platform is located, the types of users they have}
% summarize and point to Table 2 of appendix. 
%\todo{More precision with regards to what we deem user-generated content?}

%% file: sections/method.tex
\begin{figure*}
    \includegraphics[width=0.725\textwidth]{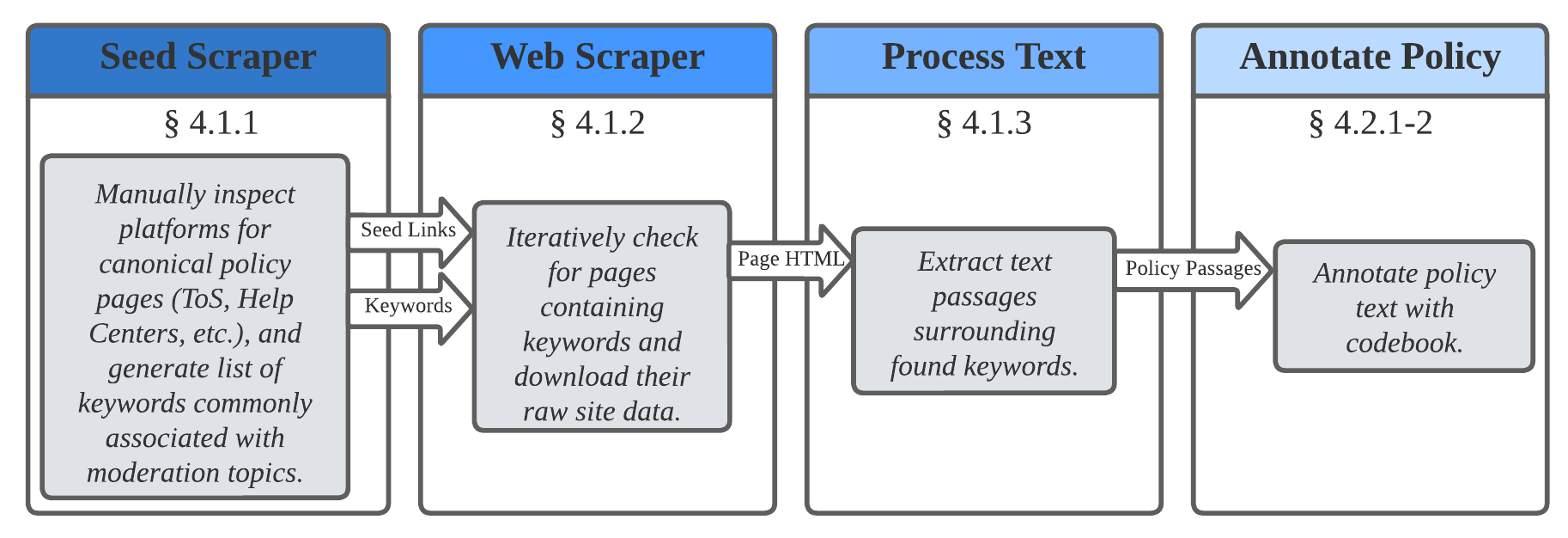}
    \caption{Pipeline for coding and annotating content moderation policy dataset OCMP-43.}
    \label{fig:pipeline}
    \Description{A flowchart describing each stage of the data pipeline for our work. There are four main stages. The first stage (leftmost box) is the Seed Scraper discussed in 4.1.1, followed by the Web Scraper discussed in 4.1.2. Then there is the Process Text stage discussed in 4.1.3. Finally, the last stage (rightmost box) is discussed Sections 4.2.1 and 4.2.2.}
\end{figure*}

In this section, we outline our procedure for obtaining relevant content
moderation policy text from the $43$ websites and three policy areas we
detailed in \cref{sec:frame}. We then describe the framework for discussing
critical components of moderation policies and how we use this framework as a codebook to
annotate the dataset. 
Figure~\ref{fig:pipeline} summarizes the pipeline that we describe in this section.

\subsection{Dataset creation} \label{subsec:corpus_creation} 

This section describes the process for building the dataset of policy text,
including the design and implementation of our custom web scraper.

% Please add the following required packages to your document preamble:
% \usepackage[normalem]{ulem}
% \useunder{\uline}{\ul}{}

\noindent \subsubsection{Locating the policies} Modern web platforms, especially the
most popular sites in our list, lack a consistent structure for distributing
information regarding content moderation policies across a site's
pages. 
Although almost all the platforms do feature a \textit{Terms of Service (ToS)} page
that outline key policies, including those regarding content moderation, many
sites also have separate \textit{Community Guidelines/Standards}, \textit{Help Centers}, and/or
official blog posts. Moreover, a particular platform's policies may be hosted on
another domain entirely---often on parent platforms or customer service
platforms. For instance, many of \url{facebook.com}'s policy pages can be found
on \url{meta.com}, and \url{reddit.com}'s policies are on both
\url{redditinc.com} and \url{reddithelp.com}. We first manually explored the
43 websites of focus and recorded URLs for each site's key policy pages. In most
cases, we found the URLs corresponding to the \textit{ToS}, as well as the \textit{Community
Guidelines/Standards} and \textit{Help Center} if they existed---we refer to these as
\emph{canonical links}.

Informed by the manual exploration, we then curated a list of keywords that
were closely associated with the three policy topics of interest (see
Table~\ref{tab:keywords}), which we refer to as the \emph{topic-wise keyword
list} henceforth.  We searched for these keywords paired with platform names
on generic search engines such as \url{google.com}, as well as on the
platforms themselves, if they had a \emph{Help Center} or other searchable database
of links. The aim of this process was to ensure we included URLs from diverse
parts of each platform such that our subsequent scraping would be able to find
all relevant policy text. The links obtained with this process, along with the
canonical links, formed the set of \textit{seed links} for the web scraper,
which
resulted in an average of 17.8 links per website, approximately evenly
distributed across the three topics. For the full list of the seed links we used, see \S A of the Supplementary Materials. 

%We host a human-readable version of the seed links at \todo{\url{anonymous-wiki.com}}.

%The keywords guided the process of  the purpose of indicating whether any given policy page contained text related to our topics and, thus, whether our scraper should download its contents. 
%To check whether these pages contained policy
%pertaining to a given content moderation topic of interest (e.g., misleading content) and could thus function as a seed link for the
%combination of that content moderation topic of interest and platform, 

\begin{table*}[]
  \caption{The keywords we associated with each content moderation topic. }
  \label{tab:keywords}
    \begin{tabular}{|c|c|c|}
    \toprule
    \textbf{Copyright Infringement} &
      \textbf{Misleading Content} &
      \textbf{Harmful Speech} 
       \\ \midrule
    {[}`copyright', `dmca'{]} &
      \begin{tabular}[c]{@{}c@{}}{[}`misinfo', `mislead', `disinfo',  `authentic', `trust',\\ `integrity', `misrepresent', `impersonat', `manipulat', \\ `decept', `deceive', `spam', `fraud', `fake', `false'{]}\end{tabular} &
      {[}`hate', `abus', `violen', `discrimin'{]} 
       \\ \bottomrule
    \end{tabular}
\end{table*}

\noindent \subsubsection{Scraping the policies} We use the set of seed links as the
starting point to explore all web pages on a given platform that may contain topic-related content moderation policies for
that platform. We designed and implemented a custom web scraper, as we found existing open-source
solutions such as Scrapy~\cite{scrapy} and MAXQDA's~\cite{maxqda} built-in
scraper often failed to retrieve text from pages of interest due to bot
blocking, dynamic loading of webpages using JavaScript, as well as direct rate
limiting for queries from a particular Internet Protocol (IP) address. Our custom scraper, which is
open-sourced along with this paper\footnote{Code Repository: \href{https://anonymous.4open.science/r/content-moderation-7AC0/}{https://anonymous.4open.science/r/content-moderation-7AC0/}}, works as follows:

\begin{enumerate}
    \item We use a headless Selenium browser using the
    \textsf{undetected-chromedriver} (with edits\footnote{We had to locally edit this package's contents to work with our version of Chrome and to work with python multiprocessing.}~\cite{ud-chrome}) to download the HTML page
    source for each of the seed links. This driver helps bypass standard
    bot-blocking techniques. 
    \item  For each HTML page, we parse the HTML to identify embedded links. If the links lead to a
    non-empty page, we check if any of the text on the page matches a keyword 	in the topic-wise keyword list. If it does, we batch the page for further
    exploration, as well as download its HTML for analysis. Page sources are stored separated by which moderation topic they relate.
    \item For each batched page, we repeat steps (1) and (2) until we have
    visited every page two hops from the seed links, scraping those that
    have any text matching the topic-wise keyword list. Note that the dataset is entirely text, meaning we do not retain images, interactive layouts, bullet points, or other non-text elements.
\end{enumerate}

The scraper can systematically locate and scrape more policy pages than is
possible from manual scraping alone. In fact, early deployments of the scraper
included more pages than necessary, often traversing URLs that were not on the
original platform at all.  To address this issue, we created an allow-list and
block-list for each platform which matches URL patterns for desired web pages,
and omits webpages that are clearly unrelated to the platform in question. In
addition, due to the unrestricted nature of webpage design, and our specific
requirements, we did not follow all links we found from a given page. For
instance, we excluded \textsf{mailto}, fragment, and anchor links. We also had
to join relative links, and follow redirection links. As per our earlier study
design choice, we did not include pages that were not in English which we
detected using \textsf{langdetect}'s pretrained models~\cite{langdetect}. A
more detailed description of our iterative scraper design in response to
issues we encountered is in \S A of the Supplementary Materials.

We conducted measurements from May to July 2023, resulting in text from 8514
policy pages from all 43 platforms. In some cases, in spite of our best
efforts at trying to bypass rate limiting, the scraper was unable to download
the page source from relevant pages. We manually copied the text from these
115 pages, forming 1.4\% of the total number of pages we scraped. 

\noindent \subsubsection{Extracting policy text} Even on the pages for which we
identified as having content moderation policy keywords, the resulting data
contains a wealth of text that is irrelevant to content moderation. This is
byproduct of Step (2) in the method, which only requires a single word on the
page to match the topic-wise keyword list for us to collect the corresponding
text. There could thus be a host of irrelevant text on the page, adding overhead
to the annotation process. For example, a \textit{ToS} page may have large
amounts of text reserved for subscriptions and fees, or text scraped from a
\textit{Help Center} post may include menu headers. To restrict the dataset to
include only relevant policy text, we aim to extract only the passages from a
page that contain the keywords. To this end, we parsed the raw text into sentences and included the 5 sentences before and after the sentence containing the keyword, merging overlapping passages as needed. Each passage is then labeled along two axes: which platform the text
came from, and which topic of content moderation the text referred
to---Copyright Infringement, Harmful Speech, or Misleading
Content.

\noindent \subsubsection{Ethical considerations around scraping} We argue here
that our scraping method abides by research ethics,
even though the Terms of Service of most the platforms we consider do not permit
scraping. First, the ruling of the United States District Court for the District
of Columbia in the case of \emph{Sandvig vs. Sessions} established
\cite{sandvig18} that it is legally permissible for researchers to use automated
tools to collect information from websites, in spite of the provision of the
Computer Fraud and Abuse Act (CFAA). Second, we do not scrape text from any
pages that contain the personal data of users. In fact, for most platforms, we
considered only pages that were accessible without logging in to the platform,
which is clearly publicly available information.

\subsection{Annotation Scheme for Dataset Analysis} \label{subsec:codebook}

\begin{table*}
\centering
\caption{The policy annotation scheme. Subcodes (where applicable) are listed below top-level codes.}
\label{tab:codebook}
\begin{tabular}{>{\raggedright}p{156pt}|>{\raggedright}p{300pt}}
\toprule
\textbf{Code}                                                            & \textbf{Memo}                                                                                                                                             \cr
\midrule
\textsc{POLICY JUSTIFICATION}                                            &                                                                                                                                                           \cr
\textsc{Community, Trust, \& Safety}                                    & References to community values or user trust  safety as motivation for policy.                                                                            \cr
\textsc{Legal}                                                           & References to extant legal frameworks for motivation for policy.                                                                                          \cr
\hline
\textsc{MODERATION CRITERIA}                                             &                                                                                                                                                           \cr
\textsc{Definition}                                                      & Definitions clarifying content that is not allowed.                                                                                                        \cr
\textsc{Example}                                                         & {Examples of content that is not allowed; can also be broad description of \newline content types.}                                                               \cr
\textsc{Exception}                                                       & {Explains content that is allowed with aim to delineat border-line cases or \newline explain special circumstances where otherwise violative content is allowed.} \cr\hline
\textsc{SAFEGUARDS}                                                      &                                                                                                                                                           \cr
\textsc{Active User Role}                                                & {When users play an active role in the content moderation, such as reporting \newline and flagging content.}                                                      \cr
{\textsc{Platform Detection Methods} /\newline \textsc{Prevention Initiatives}} & {When platforms employ methods to safeguard against violative content, \newline such as automated detection technology and moderator training initiatives.}       \cr\hline
\textsc{PLATFORM RESPONSE}                                               &                                                                                                                                                           \cr
\textsc{User-Targeted Enforcement}                                       & {Responses to becoming aware of violative content that focus on the user that \newline posted the content.}                                                       \cr
\textsc{Content-Targeted Enforcement}                                    & {Responses to becoming aware of violative content that focus on the content \newline itself.}                                                                     \cr
\textsc{Investigation / Review}                                          & {Responding to potentially violative content by investigating context or \newline gathering more information.}                                                    \cr\hline
\textsc{REDRESS / APPEAL}                                                & User pursuit of an enforcement being reconsidered/overturned.                                                                                              \cr
\hline
\textsc{BINDING LEGALESE}                                                &                                                                                                                                                           \cr
\textsc{Liability}                                                       & {Platform explains their lack of liability for actions related to content \newline moderation policy (non-)enforcement.}                                          \cr
\textsc{User Rights Altered}                                             & {Policy text that calls out the altering of user rights related to content\newline moderation; often includes phrases such as ``you warrant/agree.''}   \cr\hline
\textsc{SIGNPOST}                                                        & {When platform policy links to information on another page such as other\newline policy pages or third-party resources.} \cr
\bottomrule                                         
\end{tabular}
\end{table*}

To provide further structure to the large volume of text gathered from the
content moderation policies of the platforms we consider, we devised an
annotation scheme (or ``codebook'') to label relevant sections of text.
Such an approach is commonly used in the social sciences to both categorize
and extract meaning from language-based data~\cite{saldana2021coding}. As is
common, we combined both deductive and inductive approaches to codebook
development. Using the deductive approach, we developed an initial list of
codes that could help answer our hypotheses with regard to the composition of
content moderation policies, and their link to existing legal regimes. We
focused on codes that could capture the \emph{purpose} of a portion of policy
text, especially as it pertained to how a user would interact with it. We were
guided by the principles outlined in Kiesler et
al.~\cite{kiesler2011regulating} for healthy online communities, with
emphasis on ``moderation criteria, a chance to argue one's case, and appeal
procedures'' as key components of a comprehensive content moderation policy.
The team, which includes a legal expert on free speech laws, iteratively
refined the codebook over multiple cycles of coding subsets of the text
corpus. The codebook thus both informs the analysis of the text, and is
informed by the text itself. The codebook serves the dual purpose as both an
annotation schema and a framework of critical components for content moderation
policies. The complete codebook is presented in Table~\ref{tab:codebook}. The
codebook makes use of both high-level categories for the codes (e.g.,
`Policy Justification' and `Platform Response') and more specific sub-codes (e.g.,
`Policy Justification > Legal' and `Platform Response > User-Targeted Enforcement').

\begin{figure*}
    \centering
    \includegraphics[width=0.8\textwidth]{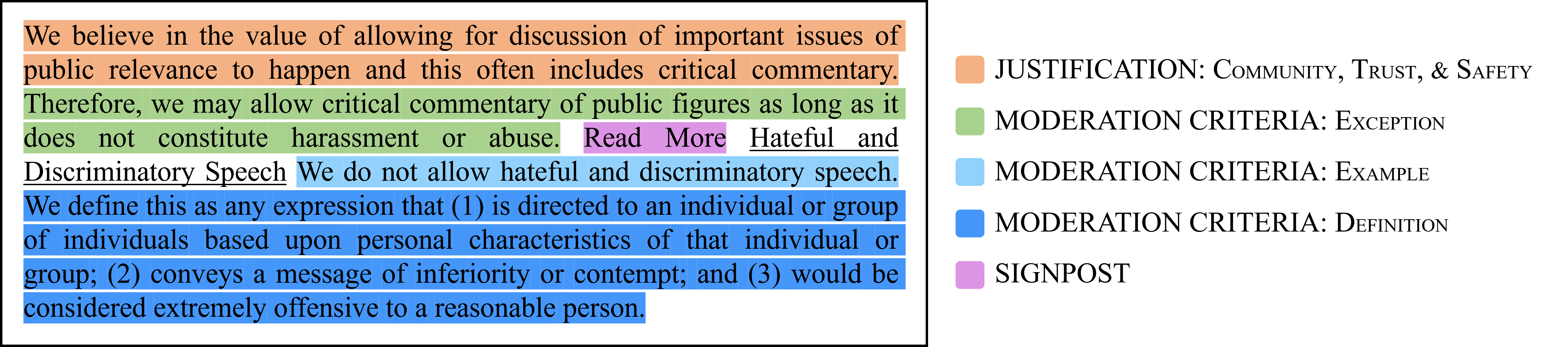}
    \caption{An example of an annotated policy passage.}
    \label{fig:exampleannotated}
    \Description{Included is a screenshot of a passage of policy text. We highlighted different parts of the passage corresponding to a code from our policy annotation scheme. The text ``We believe in the value of allowing for discussion of important issues of public relevance to happen and this often includes critical commentary'' is labeled with the code Justification: Community, Trust, \& Safety. Following, the text ``Therefore, we may allow critical commentary of public figures as long as it does not constitute harassment or abuse'' is labeled with Moderation Criteria: Exceptions. Then, the text ``Read More'' is coded with Signpost. Then, there is a policy heading, uncoded, that says Hateful and Discriminatory Speech. Then, the text ``We do not allow hateful and discriminatory speech'' is coded with Moderation Criteria: Example. Finally, the text ``We define this as an expression that (1) is directed to an individual or group of individuals based upon personal characteristics of that individual or group; (2) conveys a message of inferiority of or contempt; and (3) would be considered extremely offensive to a reasonable person'' is all coded with Moderation Criteria: Definition. }
\end{figure*}

Four members of the research team were involved in iterative codebook development and coding process. For each platform, one coder
annotated all policy-page text with the final codebook in
MAXQDA. The first coder
annotated all the policy pages for the site (across all three topics of interest \textit{copyright,
harmful speech}, and \textit{misleading content}) so that they could understand
the required context of a platform's site-wide policy, such as self-referenced
pages and guidelines. A secondary coder then checked through the annotated policies,
indicating spots for further discussion, which we discussed and resolved in many
recurring research meetings. Each coder performed both primary and secondary
coding split across different platforms, dispersing perspective and accountability around the dataset. An example of a coded policy passage is shown in Figure~\ref{fig:exampleannotated}.
% When applying codes to policy text, coders maintained context awareness--- \todo{Say we were context-aware while coding} I think it might actuall make it more confusiing if we explain that.

The policy annotation enables powerful mixed-methods analysis (see
\cref{sec:analysis}). First, it allows us to effectively locate instances of
policy text across the different codes, which we rely on as evidence in support
of hypotheses regarding the policies. Second, code coverage can indicate
platforms' focus on a given topic, as well as differences across platforms. Our
final codebook serves as an initial taxonomy for the critical components of
content moderation policy. For example, `Moderation Criteria' refers to
\textbf{What} is moderated (and what may not be); `Policy Justification' refers to
\textbf{Why} content is moderated; `Safeguards', `Platform Response', and
`Redress / Appeal' are different temporal aspects of \textbf{How} the process of
content moderation manifests. Further, `Binding Legalese' provides information
on \textbf{Whom} liability and responsible falls for content, while `Signposts'
describes the structural components of the policy text.

\subsection{Methodological Challenges and Limitations}\label{subsec:limitations}
With any data collection and annotation process of this scale, there are a
number of challenges we faced, and limitations with respect to the final
dataset. We document these in this section so users of the collection pipeline as well as
the dataset can adjust future usage and analysis accordingly.

\subsubsection{Data collection} Our scraper implementation introduces several
limitations that may affect this work's conclusions. First, relying on keyword for corpus
creation introduces the possibilities of a dataset not perfectly representative
of extant content moderation policies. To combat this issue, we removed false
positives (unrelated policy text) when applying qualitative codes, and we added
the iterative scraper to reduce false negatives (uncaptured policy text). Still,
there may be auxiliary policy text that was inaccessible to our scraper and
therefore
not represented in our dataset. Second, if a platform changed its policy pages
during the study, we cannot guarantee that our seed links or allow/block-lists
are up-to-date. Third, we discovered that a substantial fraction of the extracted text
files contained no relevant policy due to the ubiquity of terms such
``copyright'' and ``safety'' in the headers and footers, which triggered our
keyword matching. However, due to the unstructured nature of HTML and
inconsistent choices across different platforms (and even pages within the same
platform), we found no clear way to distinguish portions of a webpage that
comprised the body. We thus left this filtering to the (manual) annotation step, which
increased the load on the coders. Finally, we ended up with number of pages that
had duplicated content but different URLs. We relied on the annotation process
to exclude these from the analysis, but 
different design goals could change the trade-off between the reliance on the
scraping versus the annotation process.

\subsubsection{Data annotation and analysis} \label{analysislimitations} First, the annotation scheme was developed on a
relatively small number of platforms in comparison to all possible of platforms
that host user-generated content. However, these platforms still generated a
large amount of policy text and taken together, represent a large portion of the
Internet's user-generated content. We are thus confident that the annotation
scheme will remain useful for the platforms that host user-generated
content. Second, although we took care to disambiguate the labels in our
annotation scheme as much as possible from each other, each coder still had to
make choices about certain labels that may be close. In particular, deciding
when policy text referencing moderation criteria was an \textsc{Example} and when it
functioned as a \textsc{Definition} was challenging. We often saw the two interleaved,
with definitions bleeding into examples and \emph{vice versa}. We erred on the
side of marking moderation criteria largely as examples unless they specifically
included phrases such as ``we define''. Our choice here directly impacts
Finding~\ref{cfind: definitions}. 
Third, it is noteworthy that a platform's stated policies may differ 
from their actual moderation practices. This study focuses on what platforms 
communicate rather than on their operational conduct. While we acknowledge 
potential gaps between policy and action, we view this as a motivation for 
our work. Recognizing what platforms express they do is fundamental to grasping
their operational reality.

Finally, the process of categorizing platforms for analysis poses significant challenges,
as platforms often possess multiple facets that defy straightforward classification. 
% The nature of leading website categorization tools illustrates this complexity: Cloudflare Radar 
%, a well-studied modern challenge~\cite{}. 
% Relevant popular tools like Alexa's deprecated categorization system and Cloudflare's
% overlapping category approach emphasize the complexity of this task. 
Consequently, we resist analysis based on platform categorizations due to this
inherent difficulty, apart from one exception where we analyze the completeness
of policy components for select platform categories---specifically, those where
we can clearly delineate closed, exhaustive groups from among the 43 platforms.
We concentrate on three categories: (1) \emph{Adult content}: \url{pornhub.com},
\url{xhamster.com}, \url{xnxx.com}, \url{spankbang.com}, and \url{xvideos.com};
(2) \emph{News}: \url{nytimes.com}, \url{theguardian.com},
\url{bbc.co.uk}, and \url{washingtonpost.com}; (3) \emph{E-commerce}:
\url{amazon.com}, \url{ebay.com}, \url{etsy.com}, and \url{aliexpress.com}.

% We use comparative quantitative analysis to determine when
% policies do not contain certain components that other policies contain or reveal
% where policies asymmetrically emphasize certain critical components over others.
% This applies for both across platforms and across content moderation areas. For
% instance, Platform A may provide legal justification for their copyright
% moderation policy where Platform B may recite their copyright moderation policy
% without legal justification. Secondly, platforms may be more likely to contain
% policy text containing each critical part of the taxonomy for one moderation
% area over another, indicating more well-defined and layed-out policies for
% certain moderation areas over others. We then conduct follow-up qualitative
% analysis, providing context and reasoning about why certain comparative
% conclusions may exist with thematic analysis of the relevant coded portions. 

%\noindent \textbf{Iterative creation of a taxonomy for components of policy text} \\
%\emph{Why?} 1. Needed to understand what was in the policies, so that we could analyze variations among policies; 2.Reason about why certain text is included; 3. Acts as a guideline for what policies should contain, and determine when certain policies lack key components; 4. Terms in the taxonomy are used to code our corpus for quantitative and qualitative analysis

%% file: sections/dataset.tex
In this section, we provide an overview of the annotated dataset (OCMP-$43$)
generated using the method that we described in \cref{sec:method}. We provide
statistics about the composition of the annotated dataset, as well as where the
policy text was located on respective pages.

\begin{table*}[]
\caption{Descriptive Metrics for the Dataset OCMP-43}
\label{tab: dataset_descriptors}
\begin{tabular}{l|r|r|r|r}
\toprule
\textbf{Descriptive Metric} &
  \multicolumn{1}{c|}{\textbf{\begin{tabular}[c]{@{}c@{}}Copyright\\ Infringement\end{tabular}}} &
  \multicolumn{1}{c|}{\textbf{\begin{tabular}[c]{@{}c@{}}Harmful\\ Speech\end{tabular}}} &
  \multicolumn{1}{c|}{\textbf{\begin{tabular}[c]{@{}c@{}}Misformation\\ \end{tabular}}} &
  \multicolumn{1}{c}{\textbf{Total}} \\ \midrule
No. Coded Segments     & 3,953              & 3,034              & 4,374
    & 11,361                \\ \hline
Coded/Total Pages      & 390/2,739          & 342/1,546          & 570/4,229
    & 1,302/8,514            \\ \hline
Coded/Total Characters & 475,631/8,275,886 & 401,294/5,580,974 & 584,525/19,671,028 & 1,461,450/33,527,888 \\ \bottomrule
\end{tabular}
\end{table*}

\subsection{Dataset descriptors} We create a repository of text files
organized first by platform, and subsequently by content topic for each
website from a scraper run at a specific point in time. Each text file
contains policy text from a single page, organized by passages within. It also
indicates the URL of the page to enable reproducibility.  Each passage is
labeled with the keyword from the topic-wise keyword list that led to that
passage being included, which in turn led to the page being included. We note
that each file can contain passages with policy text corresponding to
different keywords. During the annotation process, we noticed that the same
policy text was sometimes repeated across different pages.  We used  our
discretion in flagging these to be excluded in any analysis.

In Table~\ref{tab: dataset_descriptors}, we provide an overview of the annotated
dataset. There is a clear difference in coding coverage across the three topics,
with harmful speech having the largest fraction both in terms of pages coded
($21.7\%$) as well as characters coded ($7\%$). This occurs due to the frequent
presence of terms like ``copyright'' and ``spam'' or ``trust'' in the headers
and footers of webpages, which leads to our scraper acquiring pages that 
were not coded. \S B of the Supplementary Materials contains a more detailed
platform-wise breakdown of the metrics from Table~\ref{tab: dataset_descriptors}.

\noindent \textbf{Computation and memory requirements:} The complete dataset
comprises 8,514 text files, taking about 35 MB uncompressed. We ran the
scraper and extractor on a Intel Xeon server equipped with a AMD EPYC 7502P
32-Core Processor. With our default settings, the process of scraping and
extracting the complete dataset took about 86 compute hours (which we reduced
with multiprocessing).

\subsection{Where is Policy Text Located?} \label{sec:location}

From a user's perspective, it is important to be able to locate policy
concerning content that they think may be subject to moderation. They may be checking to determine
if they are likely to face enforcement for content they wish to post, be
looking for measures to flag posts from another user, or seeking recourse if they feel they have been unfairly moderated. In light of these user needs, we
analyze where policy regarding content moderation is located on different
platforms and how straightforward it is to navigate to the policy from the
landing page.

Content moderation policies are typically found in three different
places across different platforms: the \emph{Terms of Service}, a \emph{Help
Center}, and an additional page (or set of pages) that broadly defines how
users should conduct themselves on the platform. For brevity, we refer to the
latter as \emph{Community Guidelines}, although different platforms term these
differently, with variations like `community rules', `code of conduct' and so
forth.

As we expected to some degree, all but one of 43 platforms we consider has a ToS page,
which in all cases is linked directly from the landing page of the platform. The
World Wide Web (W3) Consortium's website is the only one that does not link to a
ToS. The ToS contain broad instructions and information on the
conditions under which users will use the platform and its services. It also
contains what we call \texttt{Binding Legalese}
in our codebook (see Table~\ref{tab:codebook}), which is text that fundamentally
alters users' rights when using the platform. While a majority of platforms have
ToS accessible directly from the landing page, this is not true for the
\emph{Community Guidelines}.

\noindent \textbf{Location Observation 1:} $79\%$ ($34/43$) of platforms have page(s)
dedicated to \emph{Community Guidelines}, where platforms lay out their
expectations of user behavior, especially as pertaining to inter-user
interaction. However, only $35\%$ ($12/34$) of these platforms link to the \emph{Community
Guidelines} from the landing page. During the process of collecting the dataset, we added the \emph{Community Guidelines}
as a canonical link by explicitly looking for them via external search engines.
We were driven to do this to be exhaustive as motivated by our
research questions, but this is not the approach most users have when
interacting with a platform. We thus find it significant that only about a third of
the platforms we consider link to the \emph{Community Guidelines} directly from the
landing page. This observation raises interesting questions for future user studies
to determine the extent to which users are aware of the presence of \emph{Community
Guideline} and consciously hew to them when interacting on the platform.

\noindent \textbf{Location Observation 2:} $84\%$ ($36/43$) of platforms have a
\emph{Help Center} that directs users to policy relevant to content moderation,
with $97\%$ ($35/36$) of these help centers linked from the landing page. In
addition to the $36$ platforms with \emph{Help Centers} that contain text
relevant to content moderation policy, three others have help centers but do not
contain any relevant policy. Interestingly, one platform, \url{quora.com}, required
users to be logged in to access both the \emph{Community Guidelines} and
\emph{Help Center} from the landing page. From a user perspective, it is
encouraging that most large platforms have an easily accessible help center.
However, these only lead to relevant policy when using terms from a topic-wise
keyword list, which users may not think of when searching for these pages.

\noindent \textbf{Location Observation 3:} Content moderation policy is found scattered
across pages including, but not limited to, \emph{Transparency Centers}, privacy,
advertising, developer, and seller policies, and blog posts. Platforms have a diversity of user types based on their purpose, such as
advertisers, sellers, journalists, researchers, and property owners; with each
type of user expected by the platform to conform to certain behavioral norms. We
find that this diversity correlates with the profusion of locations on a platform where
content moderation policy may be located. In addition, platforms can announce
new policies and initiatives via internal newsrooms or blogs. Although the intent
of the scattered policies is often similar, the language and organization
differs based on the user type targeted.

%% file: sections/analysis.tex
In this section, we present our findings from analysis of the content moderation policy corpus
(\cref{subsec:corpus_creation}). We use the codebook developed in
\cref{subsec:codebook} to analyze the text from the corpus both qualitatively
and quantitatively. 
Our complete annotated dataset is large, containing thousands of pages
and tens of millions of characters. We present the following findings by introducing a question followed by findings of statistical trends, for which we employ percentages of coding frequencies.\footnote{We also analyzed trends using code coverage rather than code frequencies, calculating percents using length of coded text instead of number of occurrences of a code. Since all trends were consistent between both metrics, we opt to present code frequency. See \S B of the Supplementary Materials for detailed tables and equations used for each statistic reported as well as alternative methods to calculate relevant statistics.} 
We then support the statistical trends with representative quotes from the dataset of policy text. Note all emphasis is our own.

\subsection{Why do platforms say they moderate?}
Our initial analysis of the policy text revealed that platforms largely cite two
reasons for content moderation: meeting legal requirements (\textsc{Legal}) and maintaining
community standards (\textsc{Community, Trust, \& Safety}). We study the variation of these reasons across the three
moderation topics.

\begin{froval}
    \begin{cfinding}\label{cfind: justification} Of all the policy text that
    cites meeting legal requirements as the justification for moderation,
    $73.5\%$ is for copyright infringement, with only $13.8\%$ for harmful speech and $12.7\%$ for
    misinformation and misleading content.
    \end{cfinding}
\end{froval}

We hypothesize that the presence of strong legal regimes for copyright
infringement in the U.S. (e.g., DMCA) as well as other countries leads to legal requirements
being cited more commonly under the topic of copyright infringement, as shown in this exemplary text from \url{microsoft.com}: 

\begin{quote}
    ``However, \textbf{we are generally required by law} to disable access to copyrighted content (including videos, music, photographs, or other content you upload onto a Microsoft website) if the copyright holder claims that the use of the copyrighted work is infringing.'' (microsoft.com's Copyright FAQ)
\end{quote}

Laws such as ``NetzDG'' from Germany and the US Federal Trade Commission (FTC) regulations on false advertising likely account for the presence of legal justification for moderation in the other topics. The following policy text represents a typical example from the dataset that captures both major reasons platforms moderate (legal requirements and community values):
% \begin{quote}
%     ``The Network Enforcement Act (``NetzDG'') is a German law that requires
%     social networks to maintain a procedure for handling complaints about
%     unlawful content. We take a two-step approach to reviewing content that is
%     reported through the NetzDG reporting form. First, we review the reported
%     content under our Community Guidelines. If content that is reported to us
%     through the NetzDG reporting form violates our Community Guidelines, it's
%     removed from Instagram globally and the review process concludes. Second, if
%     the reported content doesn't violate our Community Guidelines, we review it
%     for legality based on the report. In this part of the process, an assessment
%     is made of whether the reported content violates the relevant provisions of
%     the German Criminal Code listed in the NetzDG.'' (\textsf{instagram.com}'s
%     Help Center)
% \end{quote}

\begin{quote}
    ``We take a two-step approach to reviewing content that is
    reported through the NetzDG reporting form. First, \textbf{we review the reported content under our Community Guidelines.} [...] Second, if    the reported content doesn't violate our Community Guidelines, [...] \textbf{we review it for legality} based on the report.'' (\textsf{instagram.com}'s Help Center)
\end{quote}

Across all three topics, it is insightful to compare how often the justification is
that the platform needs to maintain community standards.

\begin{froval}
    \begin{cfinding}\label{cfind: justification2}   
For harmful speech and misleading content, platforms reference \textsc{Community,~Trust,~\&~Safety} as their motivation for moderation $83.2\%$ and
    $82.0\%$ of the time, respectively, compared to just
    $21.3\%$ of the time for copyright infringement.\end{cfinding}
\end{froval}

% To address why they have community standards that lead them to moderate harmful
% speech, for example, in the first place, platforms can have pithy
% justifications as exemplified by this extract representative of others in the dataset:
When platforms are not guided by comprehensive \textsc{Legal} structures (such as the case with most harmful speech and misleading content), they rely on their \textsc{Values,~Trust,~\&~Safety}, exemplified by Tumblr's following pointed policy justification:

\begin{quote}
    ``Our users, as decent human beings, don't like that kind of thing.''  (Hate
    and Harassment section of \textsf{tumblr.com}'s Global Advertising Policy)
\end{quote}

% \brennan{"yes most of the
% legal justification is copyright but not the whole story" can bring up in here
% the linkdein and amazon legal stuff. }
% \brennan{also can mention ftc advertising guidelines for misleading content, and fair employment practices for discrimination. And then anything for non-legal}

% \begin{froval}
%     \begin{cfinding} 
%     $x \%$ of platforms only cite laws from
%     the United States to justify moderation. $y\%$ also cite laws from other
%     countries.
%     \end{cfinding}
% \end{froval}

\subsection{How do platforms describe what they moderate?}
Platforms need to clearly specify what is likely to get moderated, and what is
not, since this provides clarity to users, fostering healthy
communities~\cite{kiesler2011regulating}. However, across all three topics, we
found that platforms established what they were likely to moderate using
\textsc{Examples}, instead of \textsc{Definitions}.

\begin{froval}
    \begin{cfinding}\label{cfind: definitions}   
    \textsc{Definitions} only comprise $7.5\%$, $6.6\%$, and $2.3\%$ of moderation criteria in policy text regarding copyright infringement, harmful speech, and misinformation and misleading content, respectively. Describing moderation criteria by \textsc{example} accounts for the rest.
    \end{cfinding}
\end{froval}

Even when platforms did choose to include definitions for the content they moderate, it was usually accompanied by a list of clarifying examples, such as: 

\begin{quote}
    ``\textbf{Hate speech is defined as} a serious attack on a group or individual based on their race, ethnicity, gender, nationality, sexual orientation, sex, religion, caste, serious medical condition, or disability. [...] \textbf{Examples include:} - Statements like `$\langle$insert race/ethnicity$\rangle$ are not welcome in our country'.'' (quora.com's Platform Policies)
\end{quote}

% \begin{quote}
%     ``\textbf{Hate speech is defined as} a serious attack on a group or individual based on their race, ethnicity, gender, nationality, sexual orientation, sex, religion, caste, serious medical condition, or disability. This includes the use of slurs in a disparaging way, as well as content that dehumanizes or calls for violence, exclusion, or segregation of protected classes. \textbf{Examples include:} - Statements like “[insert race/ethnicity] are not welcome in our country” - Statements like “[insert religious group] are parasites” - The use of racial slurs in a disparaging way - Denial of the following events, in which protected classes were the primary targets, is prohibited: the Holocaust and Armenian Genocide.'' (quora.com's Platform Policies)
% \end{quote}

The specificity of what definitions we did find also varied by moderation topic,
with harmful speech and copyright infringement tending to be more clearly defined as shown by the previous example for harmful speech and the following for copyright infringement:

\begin{quote}
    ``\textbf{Copyright infringement is} doing any of the following without permission
    from the copyright owner(s): making copies, distributing the work (such as
    uploading to SoundCloud), performing or displaying the work publicly, or
    making `derivative works'.'' (\textsf{soundcloud.com}'s Help Center)
\end{quote}

% \begin{quote}
%     ``\textbf{Hateful conduct is} any content or activity that promotes, encourages, or
%     facilitates discrimination, denigration, objectification, harassment, or
%     violence based on the following characteristics, and is strictly prohibited:
%     Race, ethnicity, or national origin Religion Sex, Gender, or Gender Identity
%     Sexual Orientation Age Disability or Medical Condition Physical
%     Characteristics Veteran Status.'' (\textsf{twitch.com}'s Community Guidelines)
% \end{quote}

For misleading content we found most platforms adopted the following typical, rather vague tone and
lack of specificity:

\begin{quote}
    ``We define misinformation as content with a claim that is determined to be
    false by an authoritative third party.'' (\textsf{facebook.com}'s Community
    Guidelines)
\end{quote}
Examples are used either to expand on definitions in simpler, procedural
terms as exemplified by this quote:

\begin{quote}
    ``If you do perform a cover song in a live Twitch stream, please make a good
    faith effort to perform the song as written by the songwriter(s), and \textbf{create all audio elements yourself, without incorporating instrumental tracks,
    music recordings, or any other recorded elements owned by others.}''
    (\textsf{twitch.tv}'s Music Guidelines)
\end{quote}
Otherwise, in lieu of definitions entirely, platforms sometimes sought to establish concepts by example:

\begin{quote}
    ``Refrain from using broad and vague terms that have a potential to mislead
    your buyers (such as `environmentally friendly' or `eco-friendly').''
    (\textsf{etsy.com}'s Recycled Content Policy)
\end{quote}
This dichotomy between definitions and examples both being used to guide users
towards acceptable behavior on online platforms mirrors that between ``rules''
and ``standards'' when differentiating legal from illegal
behavior as stated in~\cite{klonick2017new,schafer2002legal}.

We also found surprisingly specific examples of disallowed harmful speech:

% \begin{quote}
%     ``Do not post [...] Dehumanizing speech or imagery in the form of
%     comparisons, generalizations, or unqualified behavioral statements (in written
%     or visual form) to or about: Insects (including but not limited to: cockroaches,
%     locusts) Animals in general or specific types of animals that are culturally
%     perceived as intellectually or physically inferior (including but not limited
%     to: Black people and apes or ape-like creatures; Jewish people and rats; Muslim
%     people and pigs; Mexican people and worms)'' (\textsf{facebook.com}'s Community Standards)
% \end{quote}
\begin{quote}
    ``Do not post [...] Dehumanizing speech or imagery in the form of
    comparisons, generalizations, or unqualified behavioral statements [...]
    (including but not limited to: \textbf{[censored explicit list of several harmful stereotypes]}~\footnote{We redacted the list of stereotypes to minimize potential harm to readers. 
    The original text can be found on the following page, which contains examples of stereotypes that some readers may find offensive: 
    \url{https://web.archive.org/web/20231120210834/https://transparency.fb.com/policies/community-standards/hate-speech/}}).'' (\textsf{facebook.com}'s Community Standards)
\end{quote}
The positive impact of these type of examples are
unclear~\cite{kiesler2011regulating} and could be triggering for some
users. 

Finally, platforms sometimes specified conditions under which content
that would typically be moderated is allowed. For copyright infringement, \textsc{Exceptions} mostly referenced fair use, such as: 

\begin{quote}
``\textbf{[U]sers are allowed to use copyright works} without the
authorization of the copyright holder \textbf{for quotation, criticism, review and for
the purpose of caricature, parody or pastiche provided that such use is fair}.''
(\textsf{tiktok.com}'s Intellectual Property Policy)
\end{quote}
For \textsc{Exceptions} to harmful speech, platforms considered educational or historical value:
\begin{quote}
``\textbf{We may label rather than remove content} that evokes hateful rhetoric
(including slurs) \textbf{in the context of counter speech, reclamation, or members'
personal experiences} with racism, sexism, ableism, and other forms of prejudice
or discrimination.'' (linkedin.com's Help Center)
\end{quote}
In some cases, platforms even considered newsworthiness as an \textsc{Exception} to their misleading content policies: 
\begin{quote}
``Related to the elections: \textbf{We may not take action on violating content that is
deemed newsworthy}.'' (\textsf{tiktok.com}'s Election Integrity Policy)
\end{quote}

\subsection{How do platforms find content that may need moderation?}
In order to detect and then subsequently act upon content that may need to be
moderated, we find platforms in general largely use three methods: automated detection, human moderators, and users flagging content, as summarized by this representative quote: 

\begin{quote}
    ``Pornhub moderates user-uploaded content in three major ways: through the use
    of \textbf{automated detection technologies}, through \textbf{real-life human moderators}, and
    through \textbf{user-generated reports}. '' (pornhub.com's Help Center)
\end{quote}

% \brennan{consistency of quote presentation and attribution)

We categorize the use of human moderators and automated detection technologies
together under \textsc{Platform Detection
Methods / Prevention Initiatives}, and user-generated reports under
\textsc{Active User Role} to illustrate the potential imbalances between platform and user roles.

\begin{froval}
    \begin{cfinding} \label{cfind: safeguards} Platforms rely heavily on an
        active user role when designing safeguards against content from all
        three topics that may need moderation. A large fraction of policy
        text that delineates safeguards for copyright ($83.3\%$), harmful speech
        ($61.5\%$), and misleading content ($51.0\%$) references users taking an
        active role.
        % Platforms rely heavily on user
        % involvement to detect and prevent copyright infringement with only 
        % $10.8\%$ of policy text around safeguards, referencing the platform's
        % active role, but appear to be more active for harmful speech and 
        % misleading content with $38.4\%$ and $51\%$, respectively.
    \end{cfinding}
\end{froval}

Some platforms boast successful automated approaches, as captured by this quote typifying what we saw: 
\begin{quote}
    ``While this is an ongoing journey that we continue to refine, we’re pleased that the metrics in our most recent Transparency Report show that, in the first half of 2021, close to 66.3 million violative pieces of content were removed from the site. Of these, \textbf{99.6\% were removed through our automated defenses}.'' (LinkedIn's Data Science Manager's Blog Post)
\end{quote}
These automated approaches, platforms claim, are especially successful for blocking spam, a well-studied problem since the advent of digital communication. In other domains, such as harmful speech and misleading content, automated detection methods are less viable when policy-violating language cannot be articulated and used for automatic detection. For instance:
\begin{quote}
    ``\textbf{Misinformation is different from other types of speech addressed in our Community Standards because there is no way to articulate a comprehensive list of what is prohibited.} With graphic violence or hate speech, for instance, our policies specify the speech we prohibit, and even persons who disagree with those policies can follow them. With misinformation, however, we cannot provide such a line. The world is changing constantly, and what is true one minute may not be true the next minute.'' (facebook.com's Community Standards on Misinformation)
\end{quote}

The varying types of user-generated content also prove troublesome for automated approaches. For example, \textsf{twitch.tv}, a platform featuring mostly \textit{live} video streaming and \textit{live} chat engagement, discusses their unique challenges for automation: 
\begin{quote}
    `\textbf{`Content moderation solutions that work for uploaded, video-based services do not work, or work differently, on Twitch.} Through experimentation and investment, we have learned that for Twitch user safety is best protected, and most scalable, when we employ a range of tools and processes, and when we partner with, and empower, our community members. The result is a layered approach to safety—one that combines the efforts of both Twitch (through tooling and staffing) and members of the community, working together.'' (twitch.tv's NetzDG Transparency Report)
\end{quote}

Even with advanced automated approaches for flagging potentially problematic content, human reviewers are often needed to determine whether the flagged content does in fact violate platform policy. 
Platforms do address the somewhat heavy role that users are expected to play in
content moderation, although it is unknown what user attitudes and seriousness are toward this task:
\begin{quote}
``Q: \textbf{It seems like you're asking users to do your job for you by reporting
problems. Why should we?} A: We believe that when\textbf{ all those involved in a
community} - hosts and members, creators and contributors - \textbf{feel and take
responsibility} for maintaining an appropriate and stimulating environment, the
debate itself is improved and all those involved benefit. We work hard to make
and keep the environment constructive and convivial but \textbf{we need your help to do
so.}'' (\textsf{guardian.com}'s Frequently Asked Questions)
\end{quote}

The burden on users appears very pronounced for copyright infringement, with
only $16.7\% $ of all policy text relating to platforms' active
role coming from copyright-related policy. This is likely due to the
structure of reporting practices laid out in the DMCA, where platforms rely on
user reports to take down violating content, and await counter-notices for
possible reinstatement (see~\cref{subsec:redressal}).

We also find evidence of heavy usage of human moderators, indicating that the
promises of automated content moderation have perhaps not
materialized~\cite{meta_content_mod} as demonstrated by this quote:
\begin{quote}
``24/7 Human Moderation Team: Our team of \textbf{moderators and support staff work 24
hours a day, 7 days a week to review all uploaded content for violations},
address user concerns, and remove all content that we identify or of which we
are made aware of and deem as constituting hate speech.''
(\textsf{pornhub.com}'s Hatespeech Policy)
\end{quote}

\subsection{What happens to flagged content?}
Once content has been flagged, platforms usually respond by targeting either the
content itself, targeting the user who posted the content, or starting an
investigation. Both \textsc{user-} and \textsc{content-targeted enforcement} are
prevalent across all three content moderation topics. Having some form of both content each was also standard across platforms.
% All platforms except
% \textsf{w3.org} policy text related to \textsc{content-targeted enforcement},
% and all except \textsf{bbc.com} include policy around \textsc{user-targeted
% enforcement}. 
% These approaches are thus standard.

The following list distilled from \textsf{twitter.com}'s enforcement options
illustrates roughly all the possible enforcement actions we saw across platforms
and topics: 

% \begin{quote}
% ``Below are some of the enforcement actions that we may take.\\ Tweet-level enforcement:
% \begin{itemize}
%     \item Limiting Tweet visibility: [...]
%     \item Excluding the Tweet from having ads adjacent to it: [...]
%     \item Requiring Tweet removal: [...]
%     \item Labeling a Tweet: [...]
%     \item Notice of public interest exception: [...]
% \end{itemize}

% Account-level enforcement:
% \begin{itemize}
%     \item Suspend an account: [...]
%     \item Placing an account in read-only mode: [...]
%     \item Verifying account ownership: [...]'' (twitter.com's Help Center page on range of enforcement options)
%     \end{itemize}
% \end{quote}

\begin{quote}
``Below are some of the enforcement actions that we may take. \\\textbf{Tweet-level enforcement}:
    Limiting Tweet visibility,
    Excluding the Tweet from having ads adjacent to it,
    
    Requiring Tweet removal,
    Labeling a Tweet,
    Notice of public interest exception.

\textbf{Account-level enforcement}:
    Suspend an account,
    Placing an account in read-only mode,
    Verifying account ownership'' (twitter.com's Help Center page on range of enforcement options)
\end{quote}

In addition, we also saw platforms employ a `strike policy', where first-time violators received strikes/warnings and repeat violators received escalated punishment like account termination. Strike policies were most commonly found with respect to copyright infringement, likely due to the DMCA legally requiring platforms to implement a repeat infringers policy in order to keep their copyright Safe Harbor status. 

\begin{quote}
    ``Our `3 strikes' repeat infringement policy is implemented as follows: If one or more uploads occur after receipt of notice of a first infringement, uploaders are then given 2 chances to stop uploading videos or other content infringing any third party’s copyrights. [...] \textbf{In the event that you accumulate three (3) such notices, your account will be terminated.}” (spankbang.com's 3 Strikes Policy Page)
\end{quote}

Still, platforms sometimes developed their own unique enforcement strategies. For instance, we found the following intriguing policy response by \textsf{tumblr.com} to disinformation campaigns by the Internet Research Agency
(IRA) from Russia:

\begin{quote}
    ``What we're doing in response to the interference: First, we'll be emailing
    anyone who liked, reblogged, replied to, or followed an IRA-linked account
    with the list of usernames they engaged with. Second, \textbf{we're going to start
    keeping a public record of usernames we've linked to the IRA or other
    state-sponsored disinformation campaigns.}'' (\textsf{tumblr.com}'s Official
    Staff Blog)
\end{quote}
That is, \url{tumblr.com} pursued user-targeted enforcement by both contacting suspected users as well as public ``naming-and-shaming''. 

For harmful speech and misinformation, we find a much greater prevalence of
platforms intervening via investigation instead of immediate enforcement.
\begin{froval}
    \begin{cfinding} \label{cfind: investigation} Of the policy text annotated
        as \textsc{Platform Response: Investigation}, 
        only $14.9\%$ is for
        copyright infringement, with $43.0\%$ for harmful
        speech and $42.1\%$ for misleading content. 
    \end{cfinding}
\end{froval}

We speculate this higher prevalence arises from two reasons. First, the DMCA
requires platforms to take down content flagged as infringing copyright or face liability, so they typically only check if the claim itself meets legal requirements and don't necessarily review the potentially infringing content any further:
\begin{quote}
    ``GitHub Isn't The Judge.  \textbf{GitHub exercises little discretion in this
    process other than determining whether the notices meet the minimum
    requirements of the DMCA.} It is up to the parties (and their lawyers) to
    evaluate the merit of their claims.'' (\textsf{github.com}'s DMCA Guide)
\end{quote}
Second, what constitutes harmful speech and misleading content can sometimes be nebulous, leading
platforms to hedge as encompassed by this Twitch quote:
\begin{quote}
    ``Twitch will consider a number of factors to determine the intent and
    context of any reported hateful conduct.'' (\textsf{twitch.tv}'s Community
    Guidelines)
\end{quote}

\subsection{Do users have any recourse after being moderated?}
\label{subsec:redressal} Specific instances of content moderation may be
inconsistent with a platform's stated policies~\cite{matias2022software} and user expectations, leading
users to seek rollback of moderation. However, we find that there is little
users can concretely do unless they or their content was targeted on the basis
of copyright infringement.

\begin{froval}
    \begin{cfinding} \label{cfind: redressal} Only $9.4\%$ and
       $15.3\%$ of policy text labeled with \textsc{Redress / Appeal} falls under the topics of harmful speech and misleading
       content, respectively. Appeals regarding moderation related to copyright
       infringement dominate with $75.3\%$.
    \end{cfinding}
\end{froval}

The appeals procedure for content removed due to copyright infringement is
well-established, grounded in law, and clearly laid out for users to navigate on
most platforms, with specialized copyright policy pages often in place as illustrated by a quote from Automattic (Wordpress's parent company):

% \begin{quote}
%     ``DMCA Counter-Notifications: If you are a Vimeo user who wishes \textbf{to challenge
%     the removal of materials caused by a DMCA takedown notice, you must file a
%     counter-notification containing the following:} [...] The copyright owner(s)
%     may elect to file a lawsuit against you for copyright infringement. If we do
%     not receive notice that such a lawsuit has been filed within ten (10)
%     business days after we provide notice of your counter-notification, we may
%     \textbf{restore the challenged materials.}'' (\textsf{vimeo.com}'s Copyright Policy)
% \end{quote}

\begin{quote}
    ``If you have received a Digital Millenium Copyright Act (DMCA) Infringement Notice and believe it was submitted in error, \textbf{you may submit a counter notice.} Counter notices must be submitted by the WordPress.com user who uploaded the material. \textbf{You may also use the form below to submit a DMCA counter notice:}'' (automattic.com's Online DMCA Counter Notice Form)
\end{quote}

In fact, $8.1\%$ of all policy text concerning copyright
infringement deals with user appeals, but only $1.3\%$ for
harmful speech and $1.5\%$ for misleading content does the same. This imbalance suggests that more work may need to be done to help users know what actions they can take if they feel their content has been unfairly moderated as harmful or misleading.

\subsection{How comprehensive are policies across
platforms?}\label{subsec:completeness} We find that all 43 platforms contain at
least some policy text from each of the 3 topics. While most platforms have
policy related to enforcement and expect users to play an active role in
moderation, only about half (21/43) have definitions of content moderation
criteria. To demonstrate a lacking of policy components, we \emph{define a
platform to be complete in terms of policy composition} if it contains policy
text relating to all sub-codes under Policy Justification, Moderation Criteria,
Safeguards, Platform Response and Redress/Appeal (see Table~\ref{tab:codebook}).
Binding Legalese and Signposts are often present in policies, but we do not deem
them necessary for completeness.

\begin{froval}
    \begin{cfinding} \label{cfind: completeness}
    $39.5 \% (17/43)$ of the platforms we consider are complete in terms of
    policy composition.
    \end{cfinding}
\end{froval}

As expected, many of the major platforms such as \textsf{facebook.com} and
\textsf{youtube.com} are complete, but we also find (perhaps surprisingly) 
inclusions of sites such as \textsf{pornhub.com}. 
When looking at policy completeness by platform type, we find that no specific
category of platforms has consistent completeness. 
For example \url{etsy.com} was the only complete platform out of the four e-commerce sites;
\url{pornhub.com} was the only complete platform out of the five adult content sites;
and none of the four traditional news platforms were complete.
% For example, $1/4$ e-commerce platforms, $1/4$ adult content platforms, and none of the 4 traditional news platforms are complete in terms of policy composition.
A detailed table of code-wise distribution of policy text is in \S B of the Supplementary Materials.

%% file: sections/discussion.tex
In this section, we outline the implications of our paper for different
stakeholders, as well as clear directions in which our dataset can be used and
our research methodology expanded upon.

\subsection{Implications for regulators}

The difficulty in locating, gathering, and analyzing text pertaining to
content moderation underscores a significant challenge currently facing
regulators. Specifically, although some areas of content are not subject to
moderation, certain categories (e.g., violent speech, terrorist speech,
copyrighted content) are subject to regulation, particularly in certain
jurisdictions. An important challenge regulators face is thus determining both
whether a platform's stated policy complies with laws and regulations, as well
as, ultimately, whether the platform's enforcement complies with these stated
policies.

\noindent \textbf{Standardizing policy formats and language:}
The process of gathering these policies across a large number of websites was
painstaking. We ended up finding relevant policy text across a large number of
pages for each platform, as observed in \cref{sec:location}. These pages often
had different names on different
platforms, making it difficult to determine what the intent of the page was
compared to a similar one on another platforms. This difficulty in even
locating policy text, let alone interpreting it, point to a need for dialogue
between platforms and regulators to agree upon standardization around content
moderation policy.  The current difficulties we face in even locating these
policies---regardless of the type of content---underscores the difficulty of
regulatory enforcement in this area, particularly at scale. Specifically,
while it may be feasible for a regulator to investigate a specific site
(perhaps in response to an incident), large-scale compliance checking would
require a more systematic approach. Standard locations for content moderation
policies, as well as standard (possibly even machine-readable) policies, could
make it easier for regulators to check that stated policies comply with the
laws and regulations of a particular region.
Such standardization could also ultimately help users locate relevant content
moderation policies and information on what avenues they have for recourse on
a moderation decision, e.g., in the event of what they may deem as unfair
moderation.

\noindent \textbf{Protecting consumer rights to their own content:}
In addition, we find very strongly worded text
in platforms' policies governing the right users waive when using platforms. For
example, the following excerpt is typical of the platforms' ToS. It appears extremely invasive with respect to users' right to their own content and how it may be modified and used:
\begin{quote}
    ``Specifically, you provide us with a royalty-free, irrevocable, perpetual,
    worldwide, exclusive, and fully sublicensable license to use, reproduce,
    modify, adapt, publish, translate, create derivative works from, incorporate
    into other works, distribute, perform, display, and otherwise exploit such
    content, in whole or in part in any form, media or technology now known or
    later developed.'' (\textsf{washingtonpost.com}'s Terms of Service)
\end{quote}
\noindent
We find an alarming prevalence of text that excludes platforms from any
liability for exposing users to objectionable content even as users' rights
are signed away. In addition, given the absence of 
widespread redressal mechanisms---as found in \cref{subsec:redressal}---users that disagree with any platform provisions have limited options. For instance: 
\begin{quote}
    ``\textbf{[Y]our only remedy with respect to any dissatisfaction} with (i) the Services, (ii) any term of these Terms of Use, (iii) any policy or practice of Fandom in operating the Services, or (iv) any content or information transmitted through the Services, \textbf{is to terminate your account and to discontinue use} of any and all parts of the Services.'' (fandom.com's Terms of Service)
\end{quote}
That is, often a user's only recourse upon disagreement
with any platform provisions is to stop using the platform entirely,  which can result in users migrating to other services. 

\subsection{Implications for platforms}

As difficult as it is for regulators to enforce laws concerning content
moderation, the platforms themselves might ultimately prefer to arrive at
solutions that both protect consumers while at the same time limiting the
possibility of facing potentially costly and cumbersome prescriptive content
moderation strategies.  Various areas, particularly concerning terrorist
speech and child pornography, have seen healthy industry-wide collaboration,
particularly around the sharing of datasets containing objectionable content.

Platform providers who read our paper may find the opportunity to examine
their own policies, in light of our industry-wide analysis.
Our critical analysis of existing policies and
the lens we have provided into other platforms' policy language and
practice concerning different content types provides content providers the
opportunity to improve their own policy content and structure---and perhaps
even come to a collective agreement about both the way these policies should
(or could) be structured, as well as where these policies could be placed on
their respective websites. 

Furthermore, in response to finding that most platforms in our dataset are
\textit{incomplete} in terms of expressing crucial policy components (see
\cref{subsec:completeness}), we advocate for increased transparency in
moderation policies as it directly correlates with enhancing user experience. We
argue that shedding light on the intricacies of moderation strategies,
algorithmic processes, and decision-making criteria not only builds trust with
users but also contributes to a more informed and empowered user base. That
said, we recognize the inherent tensions surrounding the implementation of
transparency. For instance, platforms may be reluctant to disclose moderation
mechanisms for fear of revealing trade secrets or aiding future moderation
evasion. This tension is apparent in the current popularity of Transparency
Reports which often include only high-level statistics on content removals and
user bans. This brings forth a complex discussion on the extent and nature of
transparency that platforms should adopt, one ripe for HCI researchers and
industry representatives to engage in.

Ultimately, such a collaborative consensus requires
not only the data and analysis that we have offered in this area, but also
a deeper understanding of consumer expectations and understanding of these
policies, and how to interpret them. Such a line of inquiry presents many
opportunities for future work in human-computer interaction research, as we
discuss in the next section.

\subsection{Implications for HCI researchers}

We suggest several avenues for future research in user and audit studies,
automation of ongoing data collection, and improvements to collection and
annotation of policy text.

\noindent \textbf{Further analysis and usage of OCMP-43:} Although our
current findings in \cref{sec:analysis} provide a comprehensive
high-level overview of why, when, and how platforms moderate, more nuanced
insights can be extracted from the dataset. For instance, future studies can
further contrast moderation policies for different types of users for the same
platforms. Platforms such as the \textsf{nytimes.com} have policy for
journalists, advertisers, and regular users, all of whom use the platform in
different ways, with some types of users being governed by legal regimes other
than the United States. Further, the policy text in OCMP-43 provides a useful
starting point for both user and audit studies in the future. User studies could
investigate how users perceive the policies of different platforms, and as well as
understand how they manage the occasionally onerous responsibilities placed on
them to moderate content themselves. Audit studies can help address the tensions around 
platform transparency discussed above by looking for
instances of policy text within OCMP-43 where platforms commit to certain
actions to moderate content and verifying if those conditions are met.
Evidence of discrepancies between a platform's stated policy and practice revealed 
through audits can strengthen demands for greater transparency. 
Our dataset greatly eases the process of finding these instances cross-platform
and by topic. 

\noindent \textbf{Using our data collection pipeline for extending OCMP-43:} Our data collection
pipeline itself (\cref{sec:method}) can be effectively used to collect additional policy text. The fact
that our scraper is built specifically to extract policy text greatly reduces
the technical burden for inter-disciplinary research. First, simply adding more
platforms and their corresponding seed links will provide researchers access to
even a larger set of online content moderation policy text. Second, a longitudinal study across
platforms can also be performed by running our scraper at regular intervals and
annotating the text that has changed from the previous run. This could provide
an insightful study of the impact of political and cultural shifts on platform
policies. For instance, there has been a large shift in the content moderation policies of \url{X.com}, formerly known as \url{Twitter} at the time we collected our data. Finally, by adjusting the topic-wise keyword list, researchers can
find policy around other topics of interest, such as data collection for
training machine learning models, or platform policies on AI generated content
being posted by users.

\noindent \textbf{Improving the collection and annotation of policy text:} As discussed in
\cref{subsec:limitations}, there are several limitations to our collection and
annotation pipelines. In the future, we would like to add further automated
checks to reduce the load on annotators having to go through irrelevant policy
pages by better parsing of HTML or even training natural language models to
identify pages that are unlikely to have relevant policy. Since annotation can
be laborious, future studies could ascertain how well a natural language model
can perform the annotation task.

%% file: sections/conclusion.tex
\section{Conclusion}

The moderation of user-generated content has
emerged as a paramount issue that must balance the needs of both user safety
and, in some countries, the right to free expression.  As the proliferation of
third-party content on these platforms continues, content moderation has
become increasingly complex due to the sheer scale of data involved, as well
as the variety of types of content that can be posted (and, ultimately, must
be moderated).  The absence of a prescriptive approach for many of these types
of content has led to challenges in consistency of content moderation policies across
different types of content, and across platforms. Lack of consistency is
apparent in both the language of the policies, and structure of the articulated policies.  This paper seeks to address these concerns with the first systematic study of content moderation
policies across 43 major online platforms, examining their variation, and
providing valuable insights for future research and policy alignment. Quantitative and qualitative analysis of the annotated dataset of policies created in this paper reveals immense variation across both topic and platform. Our data acquisition pipeline, dataset itself, and the corresponding analysis points to many future avenues for
researchers and policymakers to continue to explore, especially in terms of
how these polices might be better structured and articulated. We hope this paper acts as a catalyst for future research aimed at improving content moderation policies to enable healthier online communities.

%% file: sections/acks.tex
This work was funded by the Data \& Democracy research initiative, a collaboration between the Data Science Institute and Center for Effective Government at the University of Chicago. 

%% file: sections/appendix_equations.tex
%\subsection{Stats Equations/derivations}
\subsection{Equations for Statistical Findings}
\label{sec:percent_equations}
\label{sec:appendix_equations}

In this section, we provide the equations used to calculate the statistical findings in the main body of the paper.

\textbf{Finding 1:} The portion of segments annotated with \textsc{JUSTIFICATION > Legal} that fall into each of the three moderation areas. 

$$\frac{\textsc{Legal}\cap Copyright Infringement}{\textsc{Legal}} = 73.5\%, \frac{\textsc{Legal}\cap Harmful Speech}{\textsc{Legal}} = 13.8\%, \frac{\textsc{Legal}\cap Misleading Content}{\textsc{Legal}} = 12.7\%$$

\noindent \textbf{Finding 2:} The portion of segments regarding policy \textsc{JUSTIFICATIONS} that refer to \textsc{Community, Trust, or Safety}, split by topic.
 
$$\frac{\textsc{Community, Trust, \& Safety}\cap Copyright Infringement}{\textsc{JUSTIFICATION}\cap Copyright Infringement} = 21.3\%,$$

$$\frac{\textsc{Community, Trust, \& Safety}\cap Harmful Speech}{\textsc{JUSTIFICATION}\cap Harmful Speech} = 83.2\%,$$

$$\frac{\textsc{Community, Trust, \& Safety}\cap Misleading Content}{\textsc{JUSTIFICATION}\cap Misleading Content} = 82.0\%,$$

\noindent \textbf{Finding 3:} The portion of \textsc{MODERATION CRITERIA} segments, excluding \textsc{Exceptions}, that were annotated with \textsc{Definition}, split by topic. 

$$\frac{\textsc{Definition}\cap Copyright Infringement}{(\textsc{Definition} \cup \textsc{Example}) \cap Copyright Infringement} = 7.5\%,$$

$$\frac{\textsc{Definition}\cap Harmful Speech}{(\textsc{Definition} \cup \textsc{Example}) \cap Harmful Speech} = 6.6\%,$$

$$\frac{\textsc{Definition}\cap Misleading Content}{(\textsc{Definition} \cup \textsc{Example}) \cap Misleading Content} = 2.3\%,$$

\noindent \textbf{Finding 4:} The portion of \textsc{SAFEGUARDS} segments that were annotated with \textsc{Platform Detection Methods/Prevention Initiatives} (instead of \textsc{Active User Role}), split by topic. 

$$\frac{\textsc{Platform Detection Methods/Prevention Initiatives}\cap Copyright Infringement}{\textsc{SAFEGUARDS}\cap Copyright Infringement} = 16.7\%,$$

$$\frac{\textsc{Platform Detection Methods/Prevention Initiatives}\cap Harmful Speech}{\textsc{SAFEGUARDS}\cap Harmful Speech} = 38.5\%,$$

$$\frac{\textsc{Platform Detection Methods/Prevention Initiatives}\cap Misleading Content}{\textsc{SAFEGUARDS}\cap Misleading Content} = 49.0\%,$$

\noindent \textbf{Finding 5:} The portion of segments annotated with \textsc{PLATFORM RESPONSE > Investigation} that fall into each of the three moderation areas. 

$$\frac{\textsc{Investigation}\cap Copyright Infringement}{\textsc{Investigation}} = 14.9\%,$$

$$\frac{\textsc{Investigation}\cap Harmful Speech}{\textsc{Investigation}} = 43.0\%,$$

$$\frac{\textsc{Investigation}\cap Misleading Content}{\textsc{Investigation}} = 42.1\%,$$

\noindent \textbf{Finding 6:} The portion of segments annotated with \textsc{REDRESS / APPEAL} that fall into each of the three moderation areas. 

$$\frac{\textsc{REDRESS / APPEAL}\cap Copyright Infringement}{\textsc{REDRESS / APPEAL}} = 75.3\%,$$ 

$$\frac{\textsc{REDRESS / APPEAL}\cap Harmful Speech}{\textsc{REDRESS / APPEAL}} = 9.4\%,$$ 

$$\frac{\textsc{REDRESS / APPEAL}\cap Misleading Content}{\textsc{REDRESS / APPEAL}} = 15.3\%,$$ 

Following Finding 6, we also report the following. The portion of all coded segments within each moderation topic that discuss \textsc{REDRESS / APPEAL}.

$$\frac{\textsc{REDRESS / APPEAL}\cap Copyright Infringement}{Copyright Infringement} = 8.1\%,$$ 

$$\frac{\textsc{REDRESS / APPEAL}\cap Harmful Speech}{Harmful Speech} = 1.3\%,$$ 

$$\frac{\textsc{REDRESS / APPEAL}\cap Misleading Content}{Misleading Content} = 1.5\%,$$

\subsection{Alternative Calculations of Statistical Findings}
\label{alternative_equations}
Our analysis, described by the equations above, considers dataset-wide proportions. 
Here we also provide an alternative analysis that weights each platform equally by first calculating the relevant metric for each platform and then averaging across all 43 platforms. 
We present these alternative calculatins below, which the reader will note result in takeaways consistent with the main findings in the paper. 

\textbf{Alt. Finding 1:} The portion of segments annotated with \textsc{JUSTIFICATION > Legal} that fall into each of the three moderation areas. 

% $$\frac{\textsc{Legal}\cap Copyright Infringement}{\textsc{Legal}} = 72.3\%$$
$$AVERAGE_{i=platform}^{43} \frac{\textsc{Legal}_i \cap Copyright Infringement_i}{\textsc{Legal}_i} = 72.6\%,$$

% $$\frac{\textsc{Legal}\cap Harmful Speech}{\textsc{Legal}} = 14.4\%,$$
$$AVERAGE_{i=platform}^{43} \frac{\textsc{Legal}_i \cap Harmful Speech_i}{\textsc{Legal}_i} = 11.8\%,$$

% $$\frac{\textsc{Legal}\cap Misleading Content}{\textsc{Legal}} = 13.3\%$$
$$AVERAGE_{i=platform}^{43} \frac{\textsc{Legal}_i \cap Misleading Content_i}{\textsc{Legal}_i} = 15.6\%$$

\noindent \textbf{Alt. Finding 2:} The portion of segments regarding policy \textsc{JUSTIFICATIONS} that refer to \textsc{Community, Trust, or Safety}, split by topic.
 
% $$\frac{\textsc{Community, Trust, \& Safety}\cap Copyright Infringement}{\textsc{JUSTIFICATION}\cap Copyright Infringement} = 21.8\%,$$
$$AVERAGE_{i=platform}^{43} \frac{\textsc{Community, Trust, \& Safety}_i \cap Copyright Infringement_i}{\textsc{JUSTIFICATION}_i \cap Copyright Infringement_i} = 29.6\%,$$

% $$\frac{\textsc{Community, Trust, \& Safety}\cap Harmful Speech}{\textsc{JUSTIFICATION}\cap Harmful Speech} = 82.2\%,$$
$$AVERAGE_{i=platform}^{43} \frac{\textsc{Community, Trust, \& Safety}_i \cap Harmful Speech_i}{\textsc{JUSTIFICATION}_i \cap Harmful Speech_i} = 88.0\%,$$

% $$\frac{\textsc{Community, Trust, \& Safety}\cap Misleading Content}{\textsc{JUSTIFICATION}\cap Misleading Content} = 81.2\%,$$
$$AVERAGE_{i=platform}^{43} \frac{\textsc{Community, Trust, \& Safety}_i \cap Misleading Content_i}{\textsc{JUSTIFICATION}_i \cap Misleading Content_i} = 81.5\%$$

\noindent \textbf{Alt. Finding 3:} The portion of \textsc{MODERATION CRITERIA} segments, excluding \textsc{Exceptions}, that were annotated with \textsc{Definition}, split by topic. 

% $$\frac{\textsc{Definition}\cap Copyright Infringement}{(\textsc{Definition} \cup \textsc{Example}) \cap Copyright Infringement} = 6.9\%,$$
$$AVERAGE_{i=platform}^{43} \frac{\textsc{Definition}_i \cap Copyright Infringement_i}{(\textsc{Definition}_i \cup \textsc{Example}_i) \cap Copyright Infringement_i} = 5.7\%,$$

% $$\frac{\textsc{Definition}\cap Harmful Speech}{(\textsc{Definition} \cup \textsc{Example}) \cap Harmful Speech} = 15.2\%,$$
$$AVERAGE_{i=platform}^{43} \frac{\textsc{Definition}_i \cap Harmful Speech_i}{(\textsc{Definition}_i \cup \textsc{Example}_i) \cap Harmful Speech_i} = 2.3\%,$$

% $$\frac{\textsc{Definition}\cap Misleading Content}{(\textsc{Definition} \cup \textsc{Example}) \cap Misleading Content} = 2.5\%,$$
$$AVERAGE_{i=platform}^{43} \frac{\textsc{Definition}_i \cap Misleading Content_i}{(\textsc{Definition}_i \cup \textsc{Example}_i) \cap Misleading Content_i} = 3.9\%,$$

\noindent \textbf{Alt. Finding 4:} The portion of \textsc{SAFEGUARDS} segments that were annotated with \textsc{Platform Detection Methods/Prevention Initiatives} (instead of \textsc{Active User Role}), split by topic. 

% $$\frac{\textsc{Platform Detection Methods/Prevention Initiatives}\cap Copyright Infringement}{\textsc{SAFEGUARDS}\cap Copyright Infringement} = 10.8\%,$$
$$AVERAGE_{i=platform}^{43} \frac{\textsc{Platform Detection Methods/Prevention Initiatives}_i \cap Copyright Infringement_i}{\textsc{SAFEGUARDS}_i \cap Copyright Infringement_i} = 9.9\%,$$

% $$\frac{\textsc{Platform Detection Methods/Prevention Initiatives}\cap Harmful Speech}{\textsc{SAFEGUARDS}\cap Harmful Speech} = 38.4\%,$$
$$AVERAGE_{i=platform}^{43} \frac{\textsc{Platform Detection Methods/Prevention Initiatives}_i \cap Harmful Speech_i}{\textsc{SAFEGUARDS}_i \cap Harmful Speech_i} = 34.1\%,$$

% $$\frac{\textsc{Platform Detection Methods/Prevention Initiatives}\cap Misleading Content}{\textsc{SAFEGUARDS}\cap Misleading Content} = 51.0\%,$$
$$AVERAGE_{i=platform}^{43} \frac{\textsc{Platform Detection Methods/Prevention Initiatives}_i \cap Misleading Content_i}{\textsc{SAFEGUARDS}_i \cap Misleading Content_i} = 51.7\%,$$

\noindent \textbf{Alt. Finding 5:} The portion of segments annotated with \textsc{PLATFORM RESPONSE > Investigation} that fall into each of the three moderation areas. 

% $$\frac{\textsc{Investigation}\cap Copyright Infringement}{\textsc{PLATFORM RESPONSE}\cap Copyright Infringement} = 14.6\%,$$
$$AVERAGE_{i=platform}^{43} \frac{\textsc{Investigation}_i \cap Copyright Infringement_i}{\textsc{Investigation}_i} = 24.5\%,$$

% $$\frac{\textsc{Investigation}\cap Harmful Speech}{\textsc{PLATFORM RESPONSE}\cap Harmful Speech} = 42.2\%,$$
$$AVERAGE_{i=platform}^{43} \frac{\textsc{Investigation}_i \cap Harmful Speech_i}{\textsc{Investigation}_i} = 32.3\%,$$

% $$\frac{\textsc{Investigation}\cap Misleading Content}{\textsc{PLATFORM RESPONSE}\cap Misleading Content} = 43.2\%,$$
$$AVERAGE_{i=platform}^{43} \frac{\textsc{Investigation}_i \cap Misleading Content_i}{\textsc{Investigation}_i} = 43.2\%,$$

\noindent \textbf{Alt. Finding 6:} The portion of segments annotated with \textsc{REDRESS / APPEAL} that fall into each of the three moderation areas. 

% $$\frac{\textsc{REDRESS / APPEAL}\cap Copyright Infringement}{\textsc{REDRESS / APPEAL}} = 76.8\%,$$ 
$$AVERAGE_{i=platform}^{43} \frac{\textsc{REDRESS / APPEAL}_i \cap Copyright Infringement_i}{\textsc{REDRESS / APPEAL}_i} = 74.4\%,$$

% $$\frac{\textsc{REDRESS / APPEAL}\cap Harmful Speech}{\textsc{REDRESS / APPEAL}} = 6.9\%,$$ 
$$AVERAGE_{i=platform}^{43} \frac{\textsc{REDRESS / APPEAL}_i \cap Harmful Speech_i}{\textsc{REDRESS / APPEAL}_i} = 10.6\%,$$

% $$\frac{\textsc{REDRESS / APPEAL}\cap Misleading Content}{\textsc{REDRESS / APPEAL}} = 16.3\%,$$ 
$$AVERAGE_{i=platform}^{43} \frac{\textsc{REDRESS / APPEAL}_i \cap Misleading Content_i}{\textsc{REDRESS / APPEAL}_i} = 15.0\%,$$

Following Finding 6, we also report the following. The portion of all coded segments within each moderation topic that discuss \textsc{REDRESS / APPEAL}.

% $$\frac{\textsc{REDRESS / APPEAL}\cap Copyright Infringement}{Copyright Infringement} = 8.3\%,$$ 
$$AVERAGE_{i=platform}^{43} \frac{\textsc{REDRESS / APPEAL}_i \cap Copyright Infringement_i}{Copyright Infringement_i} = 7.3\%,$$

% $$\frac{\textsc{REDRESS / APPEAL}\cap Harmful Speech}{Harmful Speech} = 0.9\%,$$ 
$$AVERAGE_{i=platform}^{43} \frac{\textsc{REDRESS / APPEAL}_i \cap Harmful Speech_i}{Harmful Speech_i} = 0.9\%,$$

% $$\frac{\textsc{REDRESS / APPEAL}\cap Misleading Content}{Misleading Content} = 1.5\%,$$ 
$$AVERAGE_{i=platform}^{43} \frac{\textsc{REDRESS / APPEAL}_i \cap Misleading Content_i}{Misleading Content_i} = 0.9\%,$$

%% file: paper.bbl
%%% -*-BibTeX-*-
%%% Do NOT edit. File created by BibTeX with style
%%% ACM-Reference-Format-Journals [18-Jan-2012].

\begin{thebibliography}{63}

%%% ====================================================================
%%% NOTE TO THE USER: you can override these defaults by providing
%%% customized versions of any of these macros before the \bibliography
%%% command.  Each of them MUST provide its own final punctuation,
%%% except for \shownote{}, \showDOI{}, and \showURL{}.  The latter two
%%% do not use final punctuation, in order to avoid confusing it with
%%% the Web address.
%%%
%%% To suppress output of a particular field, define its macro to expand
%%% to an empty string, or better, \unskip, like this:
%%%
%%% \newcommand{\showDOI}[1]{\unskip}   % LaTeX syntax
%%%
%%% \def \showDOI #1{\unskip}           % plain TeX syntax
%%%
%%% ====================================================================

\ifx \showCODEN    \undefined \def \showCODEN     #1{\unskip}     \fi
\ifx \showDOI      \undefined \def \showDOI       #1{#1}\fi
\ifx \showISBNx    \undefined \def \showISBNx     #1{\unskip}     \fi
\ifx \showISBNxiii \undefined \def \showISBNxiii  #1{\unskip}     \fi
\ifx \showISSN     \undefined \def \showISSN      #1{\unskip}     \fi
\ifx \showLCCN     \undefined \def \showLCCN      #1{\unskip}     \fi
\ifx \shownote     \undefined \def \shownote      #1{#1}          \fi
\ifx \showarticletitle \undefined \def \showarticletitle #1{#1}   \fi
\ifx \showURL      \undefined \def \showURL       {\relax}        \fi
% The following commands are used for tagged output and should be
% invisible to TeX
\providecommand\bibfield[2]{#2}
\providecommand\bibinfo[2]{#2}
\providecommand\natexlab[1]{#1}
\providecommand\showeprint[2][]{arXiv:#2}

\bibitem[{Access Now, ACLU Foundation of Northern California}(2018)]%
        {santaclara18}
\bibfield{author}{\bibinfo{person}{{Access Now, ACLU Foundation of Northern
  California}}.} \bibinfo{year}{2018}\natexlab{}.
\newblock \bibinfo{title}{The Santa Clara Principles on Transparency and
  Accountability in Content Moderation}.
\newblock \bibinfo{howpublished}{\url{https://santaclaraprinciples.org/}}.
\newblock


\bibitem[Ahmad(2019)]%
        {ahmad_its_2019}
\bibfield{author}{\bibinfo{person}{Sabrina Ahmad}.}
  \bibinfo{year}{2019}\natexlab{}.
\newblock \bibinfo{booktitle}{\emph{"{It}'s {Just} the {Job}": {Investigating}
  the {Influence} of {Culture} in {India}'s {Commercial} {Content} {Moderation}
  {Industry}}}.
\newblock \bibinfo{type}{preprint}. \bibinfo{institution}{SocArXiv}.
\newblock
\urldef\tempurl%
\url{https://doi.org/10.31235/osf.io/hjcv2}
\showDOI{\tempurl}


\bibitem[{Amnesty}(2022)]%
        {amnesty22}
\bibfield{author}{\bibinfo{person}{{Amnesty}}.}
  \bibinfo{year}{2022}\natexlab{}.
\newblock \bibinfo{title}{Myanmar: Facebook’s systems promoted violence
  against Rohingya; Meta owes reparations}.
\newblock
  \bibinfo{howpublished}{\url{https://www.amnesty.org/en/latest/news/2022/09/myanmar-facebooks-systems-promoted-violence-against-rohingya-meta-owes-reparations-new-report/}}.
\newblock


\bibitem[Beschastnikh et~al\mbox{.}(2008)]%
        {beschastnikh2008wikipedian}
\bibfield{author}{\bibinfo{person}{Ivan Beschastnikh}, \bibinfo{person}{Travis
  Kriplean}, {and} \bibinfo{person}{David McDonald}.}
  \bibinfo{year}{2008}\natexlab{}.
\newblock \showarticletitle{Wikipedian self-governance in action: Motivating
  the policy lens}. In \bibinfo{booktitle}{\emph{Proceedings of the
  International AAAI Conference on Web and Social Media}},
  Vol.~\bibinfo{volume}{2}. \bibinfo{pages}{27--35}.
\newblock
\urldef\tempurl%
\url{https://doi.org/10.1609/icwsm.v2i1.18611}
\showURL{%
\tempurl}


\bibitem[Butler et~al\mbox{.}(2008)]%
        {butler2008don}
\bibfield{author}{\bibinfo{person}{Brian Butler}, \bibinfo{person}{Elisabeth
  Joyce}, {and} \bibinfo{person}{Jacqueline Pike}.}
  \bibinfo{year}{2008}\natexlab{}.
\newblock \showarticletitle{Don't look now, but we've created a bureaucracy:
  the nature and roles of policies and rules in wikipedia}. In
  \bibinfo{booktitle}{\emph{Proceedings of the SIGCHI conference on human
  factors in computing systems}}.
\newblock
\urldef\tempurl%
\url{https://doi.org/10.1145/1357054.1357227}
\showURL{%
\tempurl}


\bibitem[Chancellor et~al\mbox{.}(2016)]%
        {Chancellor:2016:TIC:2818048.2819963}
\bibfield{author}{\bibinfo{person}{Stevie Chancellor},
  \bibinfo{person}{Jessica~Annette Pater}, \bibinfo{person}{Trustin Clear},
  \bibinfo{person}{Eric Gilbert}, {and} \bibinfo{person}{Munmun De~Choudhury}.}
  \bibinfo{year}{2016}\natexlab{}.
\newblock \showarticletitle{\#Thyghgapp: Instagram Content Moderation and
  Lexical Variation in Pro-Eating Disorder Communities}. In
  \bibinfo{booktitle}{\emph{Proceedings of CSCW}} (San Francisco, California,
  USA) \emph{(\bibinfo{series}{CSCW '16})}. \bibinfo{publisher}{ACM},
  \bibinfo{address}{New York, NY, USA}, \bibinfo{pages}{1201--1213}.
\newblock
\showISBNx{978-1-4503-3592-8}


\bibitem[Chandrasekharan et~al\mbox{.}(2019)]%
        {chandrasekharan2019crossmod}
\bibfield{author}{\bibinfo{person}{Eshwar Chandrasekharan},
  \bibinfo{person}{Chaitrali Gandhi}, \bibinfo{person}{Matthew~Wortley
  Mustelier}, {and} \bibinfo{person}{Eric Gilbert}.}
  \bibinfo{year}{2019}\natexlab{}.
\newblock \showarticletitle{Crossmod: A cross-community learning-based system
  to assist reddit moderators}.
\newblock \bibinfo{journal}{\emph{Proceedings of the ACM on human-computer
  interaction}} \bibinfo{volume}{3}, \bibinfo{number}{CSCW}
  (\bibinfo{year}{2019}), \bibinfo{pages}{1--30}.
\newblock


\bibitem[Chandrasekharan et~al\mbox{.}(2022)]%
        {chandrasekharan_quarantined_nodate}
\bibfield{author}{\bibinfo{person}{Eshwar Chandrasekharan},
  \bibinfo{person}{Shagun Jhaver}, \bibinfo{person}{Amy Bruckman}, {and}
  \bibinfo{person}{Eric Gilbert}.} \bibinfo{year}{2022}\natexlab{}.
\newblock \showarticletitle{Quarantined! Examining the effects of a
  community-wide moderation intervention on Reddit}.
\newblock \bibinfo{journal}{\emph{ACM Transactions on Computer-Human
  Interaction (TOCHI)}} \bibinfo{volume}{29}, \bibinfo{number}{4}
  (\bibinfo{year}{2022}), \bibinfo{pages}{1--26}.
\newblock


\bibitem[Chandrasekharan et~al\mbox{.}(2017a)]%
        {chandrasekharan2017you}
\bibfield{author}{\bibinfo{person}{Eshwar Chandrasekharan},
  \bibinfo{person}{Umashanthi Pavalanathan}, \bibinfo{person}{Anirudh
  Srinivasan}, \bibinfo{person}{Adam Glynn}, \bibinfo{person}{Jacob
  Eisenstein}, {and} \bibinfo{person}{Eric Gilbert}.}
  \bibinfo{year}{2017}\natexlab{a}.
\newblock \showarticletitle{You Can't Stay Here: The Efficacy of Reddit's 2015
  Ban Examined Through Hate Speech}.
\newblock \bibinfo{journal}{\emph{In Proceedings of ACM Human Computer
  Interaction}} \bibinfo{volume}{1}, \bibinfo{number}{Computer-Supported
  Cooperative Work and Social Computing}, Article \bibinfo{articleno}{31}
  (\bibinfo{year}{2017}), \bibinfo{numpages}{22}~pages.
\newblock
\showISSN{2573-0142}


\bibitem[Chandrasekharan et~al\mbox{.}(2017b)]%
        {chandrasekharan_you_2017}
\bibfield{author}{\bibinfo{person}{Eshwar Chandrasekharan},
  \bibinfo{person}{Umashanthi Pavalanathan}, \bibinfo{person}{Anirudh
  Srinivasan}, \bibinfo{person}{Adam Glynn}, \bibinfo{person}{Jacob
  Eisenstein}, {and} \bibinfo{person}{Eric Gilbert}.}
  \bibinfo{year}{2017}\natexlab{b}.
\newblock \showarticletitle{You {Can}'t {Stay} {Here}: {The} {Efficacy} of
  {Reddit}'s 2015 {Ban} {Examined} {Through} {Hate} {Speech}}.
\newblock \bibinfo{journal}{\emph{Proceedings of the ACM on Human-Computer
  Interaction}} \bibinfo{volume}{1}, \bibinfo{number}{CSCW}
  (\bibinfo{date}{Dec.} \bibinfo{year}{2017}), \bibinfo{pages}{1--22}.
\newblock
\showISSN{2573-0142}
\urldef\tempurl%
\url{https://doi.org/10.1145/3134666}
\showDOI{\tempurl}


\bibitem[Chandrasekharan et~al\mbox{.}(2018)]%
        {chandrasekharan_internets_2018}
\bibfield{author}{\bibinfo{person}{Eshwar Chandrasekharan},
  \bibinfo{person}{Mattia Samory}, \bibinfo{person}{Shagun Jhaver},
  \bibinfo{person}{Hunter Charvat}, \bibinfo{person}{Amy Bruckman},
  \bibinfo{person}{Cliff Lampe}, \bibinfo{person}{Jacob Eisenstein}, {and}
  \bibinfo{person}{Eric Gilbert}.} \bibinfo{year}{2018}\natexlab{}.
\newblock \showarticletitle{The {Internet}'s {Hidden} {Rules}: {An} {Empirical}
  {Study} of {Reddit} {Norm} {Violations} at {Micro}, {Meso}, and {Macro}
  {Scales}}.
\newblock \bibinfo{journal}{\emph{Proceedings of the ACM on Human-Computer
  Interaction}} \bibinfo{volume}{2}, \bibinfo{number}{CSCW}
  (\bibinfo{date}{Nov.} \bibinfo{year}{2018}), \bibinfo{pages}{1--25}.
\newblock
\showISSN{2573-0142}
\urldef\tempurl%
\url{https://doi.org/10.1145/3274301}
\showDOI{\tempurl}


\bibitem[Chang and Danescu-Niculescu-Mizil(2019)]%
        {Chang-Recidivism:19}
\bibfield{author}{\bibinfo{person}{Jonathan~P. Chang} {and}
  \bibinfo{person}{Cristian Danescu-Niculescu-Mizil}.}
  \bibinfo{year}{2019}\natexlab{}.
\newblock \showarticletitle{Trajectories of Blocked Community Members:
  Redemption, Recidivism and Departure}. In
  \bibinfo{booktitle}{\emph{Proceedings of WWW}}. \bibinfo{pages}{184--195}.
\newblock


\bibitem[Chen(2023)]%
        {facebook_content_mod}
\bibfield{author}{\bibinfo{person}{Adrian Chen}.}
  \bibinfo{year}{2023}\natexlab{}.
\newblock \bibinfo{title}{{The Laborers Who Keep Dick Pics and Beheadings Out
  of Your Facebook Feed}}.
\newblock
\newblock
\urldef\tempurl%
\url{https://www.wired.com/2014/10/content-moderation/}
\showURL{%
\tempurl}


\bibitem[Cheng et~al\mbox{.}(2014)]%
        {cheng+dnm+leskovec:2014}
\bibfield{author}{\bibinfo{person}{Justin Cheng}, \bibinfo{person}{Cristian
  Danescu-Niculescu-Mizil}, {and} \bibinfo{person}{Jure Leskovec}.}
  \bibinfo{year}{2014}\natexlab{}.
\newblock \showarticletitle{How community feedback shapes user behavior}. In
  \bibinfo{booktitle}{\emph{Proceedings of ICWSM}}.
\newblock
\urldef\tempurl%
\url{https://www.aaai.org/ocs/index.php/ICWSM/ICWSM14/paper/view/8066}
\showURL{%
\tempurl}


\bibitem[Congress(1998)]%
        {congress1998digital}
\bibfield{author}{\bibinfo{person}{US Congress}.}
  \bibinfo{year}{1998}\natexlab{}.
\newblock \showarticletitle{{Digital Millennium Copyright Act}}.
\newblock \bibinfo{journal}{\emph{Public Law}} \bibinfo{volume}{105},
  \bibinfo{number}{304} (\bibinfo{year}{1998}), \bibinfo{pages}{112}.
\newblock


\bibitem[Copeland et~al\mbox{.}(1989)]%
        {Copeland:vldb1989}
\bibfield{author}{\bibinfo{person}{G. Copeland}, \bibinfo{person}{T. Keller},
  \bibinfo{person}{R. Krishnamurthy}, {and} \bibinfo{person}{M. Smith}.}
  \bibinfo{year}{1989}\natexlab{}.
\newblock \showarticletitle{The Case For Safe {RAM}}. In
  \bibinfo{booktitle}{\emph{{Proceedings of the 15th International Conference
  on Very Large Data Bases}}} (Amsterdam, The Netherlands)
  \emph{(\bibinfo{series}{VLDB '89})}. \bibinfo{publisher}{Morgan Kaufmann
  Publishers Inc.}, \bibinfo{address}{San Francisco, CA, USA},
  \bibinfo{pages}{327--335}.
\newblock
\showISBNx{1-55860-101-5}
\urldef\tempurl%
\url{http://portal.acm.org/citation.cfm?id=88830.88887}
\showURL{%
\tempurl}


\bibitem[Elliott(2023)]%
        {meta_content_mod}
\bibfield{author}{\bibinfo{person}{Vittoria Elliott}.}
  \bibinfo{year}{2023}\natexlab{}.
\newblock \bibinfo{title}{{Meta’s Gruesome Content Broke Him. Now He Wants It
  to Pay}}.
\newblock
\newblock
\urldef\tempurl%
\url{https://www.wired.com/story/meta-kenya-lawsuit-outsourcing-content-moderation/}
\showURL{%
\tempurl}


\bibitem[Fiesler et~al\mbox{.}(2017)]%
        {fiesler_what_2017}
\bibfield{author}{\bibinfo{person}{Casey Fiesler}, \bibinfo{person}{Michaelanne
  Dye}, \bibinfo{person}{Jessica~L. Feuston}, \bibinfo{person}{Chaya
  Hiruncharoenvate}, \bibinfo{person}{C.J. Hutto}, \bibinfo{person}{Shannon
  Morrison}, \bibinfo{person}{Parisa Khanipour~Roshan},
  \bibinfo{person}{Umashanthi Pavalanathan}, \bibinfo{person}{Amy~S. Bruckman},
  \bibinfo{person}{Munmun De~Choudhury}, {and} \bibinfo{person}{Eric Gilbert}.}
  \bibinfo{year}{2017}\natexlab{}.
\newblock \showarticletitle{What (or {Who}) {Is} {Public}?: {Privacy}
  {Settings} and {Social} {Media} {Content} {Sharing}}. In
  \bibinfo{booktitle}{\emph{Proceedings of the 2017 {ACM} {Conference} on
  {Computer} {Supported} {Cooperative} {Work} and {Social} {Computing}}}.
  \bibinfo{publisher}{ACM}, \bibinfo{address}{Portland Oregon USA},
  \bibinfo{pages}{567--580}.
\newblock
\showISBNx{978-1-4503-4335-0}
\urldef\tempurl%
\url{https://doi.org/10.1145/2998181.2998223}
\showDOI{\tempurl}


\bibitem[Fiesler et~al\mbox{.}(2015)]%
        {fiesler_understanding_2015}
\bibfield{author}{\bibinfo{person}{Casey Fiesler}, \bibinfo{person}{Jessica~L.
  Feuston}, {and} \bibinfo{person}{Amy~S. Bruckman}.}
  \bibinfo{year}{2015}\natexlab{}.
\newblock \showarticletitle{Understanding {Copyright} {Law} in {Online}
  {Creative} {Communities}}. In \bibinfo{booktitle}{\emph{Proceedings of the
  18th {ACM} {Conference} on {Computer} {Supported} {Cooperative} {Work} \&
  {Social} {Computing}}}. \bibinfo{publisher}{ACM}, \bibinfo{address}{Vancouver
  BC Canada}, \bibinfo{pages}{116--129}.
\newblock
\showISBNx{978-1-4503-2922-4}
\urldef\tempurl%
\url{https://doi.org/10.1145/2675133.2675234}
\showDOI{\tempurl}


\bibitem[Fiesler et~al\mbox{.}(2018)]%
        {fiesler+al:2018}
\bibfield{author}{\bibinfo{person}{Casey Fiesler}, \bibinfo{person}{Jialun
  Jiang}, \bibinfo{person}{Joshua McCann}, \bibinfo{person}{Kyle Frye}, {and}
  \bibinfo{person}{Jed Brubaker}.} \bibinfo{year}{2018}\natexlab{}.
\newblock \showarticletitle{{Reddit rules! Characterizing an Ecosystem of
  Governance}}. In \bibinfo{booktitle}{\emph{Proceedings of the International
  AAAI Conference on Web and Social Media}}, Vol.~\bibinfo{volume}{12}.
\newblock


\bibitem[Fiesler et~al\mbox{.}(2016)]%
        {fiesler_reality_2016}
\bibfield{author}{\bibinfo{person}{Casey Fiesler}, \bibinfo{person}{Cliff
  Lampe}, {and} \bibinfo{person}{Amy~S. Bruckman}.}
  \bibinfo{year}{2016}\natexlab{}.
\newblock \showarticletitle{Reality and {Perception} of {Copyright} {Terms} of
  {Service} for {Online} {Content} {Creation}}. In
  \bibinfo{booktitle}{\emph{Proceedings of the 19th {ACM} {Conference} on
  {Computer}-{Supported} {Cooperative} {Work} \& {Social} {Computing}}}.
  \bibinfo{publisher}{ACM}, \bibinfo{address}{San Francisco California USA},
  \bibinfo{pages}{1450--1461}.
\newblock
\showISBNx{978-1-4503-3592-8}
\urldef\tempurl%
\url{https://doi.org/10.1145/2818048.2819931}
\showDOI{\tempurl}


\bibitem[Fiesler et~al\mbox{.}({[n.\,d.]})]%
        {fiesler_reddit_nodate}
\bibfield{author}{\bibinfo{person}{Casey Fiesler}, \bibinfo{person}{Joshua
  McCann}, \bibinfo{person}{Kyle Frye}, {and} \bibinfo{person}{Jed~R
  Brubaker}.} \bibinfo{year}{[n.\,d.]}\natexlab{}.
\newblock \showarticletitle{Reddit {Rules}! {Characterizing} an {Ecosystem} of
  {Governance}}.
\newblock  (\bibinfo{year}{[n.\,d.]}), \bibinfo{pages}{10}.
\newblock


\bibitem[Gillespie(2018)]%
        {gillespie2018custodians}
\bibfield{author}{\bibinfo{person}{Tarleton Gillespie}.}
  \bibinfo{year}{2018}\natexlab{}.
\newblock \bibinfo{booktitle}{\emph{Custodians of the Internet}}.
\newblock \bibinfo{publisher}{Yale University Press}.
\newblock


\bibitem[Goldman(2021)]%
        {goldman2021content}
\bibfield{author}{\bibinfo{person}{Eric Goldman}.}
  \bibinfo{year}{2021}\natexlab{}.
\newblock \showarticletitle{Content Moderation Remedies}.
\newblock \bibinfo{journal}{\emph{Michigan Technology Law Review, Forthcoming}}
  (\bibinfo{year}{2021}).
\newblock


\bibitem[Gorwa et~al\mbox{.}(2020)]%
        {gorwa2020algorithmic}
\bibfield{author}{\bibinfo{person}{Robert Gorwa}, \bibinfo{person}{Reuben
  Binns}, {and} \bibinfo{person}{Christian Katzenbach}.}
  \bibinfo{year}{2020}\natexlab{}.
\newblock \showarticletitle{Algorithmic content moderation: Technical and
  political challenges in the automation of platform governance}.
\newblock \bibinfo{journal}{\emph{Big Data \& Society}} \bibinfo{volume}{7},
  \bibinfo{number}{1} (\bibinfo{year}{2020}),
  \bibinfo{pages}{2053951719897945}.
\newblock


\bibitem[Horta~Ribeiro et~al\mbox{.}(2021)]%
        {horta_ribeiro_platform_2021}
\bibfield{author}{\bibinfo{person}{Manoel Horta~Ribeiro},
  \bibinfo{person}{Shagun Jhaver}, \bibinfo{person}{Savvas Zannettou},
  \bibinfo{person}{Jeremy Blackburn}, \bibinfo{person}{Gianluca Stringhini},
  \bibinfo{person}{Emiliano De~Cristofaro}, {and} \bibinfo{person}{Robert
  West}.} \bibinfo{year}{2021}\natexlab{}.
\newblock \showarticletitle{Do {Platform} {Migrations} {Compromise} {Content}
  {Moderation}? {Evidence} from r/{The}\_Donald and r/{Incels}}.
\newblock \bibinfo{journal}{\emph{Proceedings of the ACM on Human-Computer
  Interaction}} \bibinfo{volume}{5}, \bibinfo{number}{CSCW2}
  (\bibinfo{date}{Oct.} \bibinfo{year}{2021}), \bibinfo{pages}{1--24}.
\newblock
\showISSN{2573-0142}
\urldef\tempurl%
\url{https://doi.org/10.1145/3476057}
\showDOI{\tempurl}


\bibitem[Hwang and Shaw(2022)]%
        {hwang2022rules}
\bibfield{author}{\bibinfo{person}{Sohyeon Hwang} {and} \bibinfo{person}{Aaron
  Shaw}.} \bibinfo{year}{2022}\natexlab{}.
\newblock \showarticletitle{Rules and Rule-Making in the Five Largest
  Wikipedias}. In \bibinfo{booktitle}{\emph{Proceedings of the International
  AAAI Conference on Web and Social Media}}, Vol.~\bibinfo{volume}{16}.
  \bibinfo{pages}{347--357}.
\newblock
\urldef\tempurl%
\url{https://doi.org/10.1609/icwsm.v16i1.19297}
\showURL{%
\tempurl}


\bibitem[Jhaver et~al\mbox{.}(2019a)]%
        {jhaver_did_2019}
\bibfield{author}{\bibinfo{person}{Shagun Jhaver},
  \bibinfo{person}{Darren~Scott Appling}, \bibinfo{person}{Eric Gilbert}, {and}
  \bibinfo{person}{Amy Bruckman}.} \bibinfo{year}{2019}\natexlab{a}.
\newblock \showarticletitle{"{Did} {You} {Suspect} the {Post} {Would} be
  {Removed}?": {Understanding} {User} {Reactions} to {Content} {Removals} on
  {Reddit}}.
\newblock \bibinfo{journal}{\emph{Proceedings of the ACM on Human-Computer
  Interaction}} \bibinfo{volume}{3}, \bibinfo{number}{CSCW}
  (\bibinfo{date}{Nov.} \bibinfo{year}{2019}), \bibinfo{pages}{1--33}.
\newblock
\showISSN{2573-0142}
\urldef\tempurl%
\url{https://doi.org/10.1145/3359294}
\showDOI{\tempurl}


\bibitem[Jhaver et~al\mbox{.}(2019b)]%
        {jhaver_human-machine_2019}
\bibfield{author}{\bibinfo{person}{Shagun Jhaver}, \bibinfo{person}{Iris
  Birman}, \bibinfo{person}{Eric Gilbert}, {and} \bibinfo{person}{Amy
  Bruckman}.} \bibinfo{year}{2019}\natexlab{b}.
\newblock \showarticletitle{Human-{Machine} {Collaboration} for {Content}
  {Regulation}: {The} {Case} of {Reddit} {Automoderator}}.
\newblock \bibinfo{journal}{\emph{ACM Transactions on Computer-Human
  Interaction}} \bibinfo{volume}{26}, \bibinfo{number}{5}
  (\bibinfo{date}{Sept.} \bibinfo{year}{2019}), \bibinfo{pages}{1--35}.
\newblock
\showISSN{1073-0516, 1557-7325}
\urldef\tempurl%
\url{https://doi.org/10.1145/3338243}
\showDOI{\tempurl}


\bibitem[Jhaver et~al\mbox{.}(2021)]%
        {jhaver_evaluating_2021}
\bibfield{author}{\bibinfo{person}{Shagun Jhaver}, \bibinfo{person}{Christian
  Boylston}, \bibinfo{person}{Diyi Yang}, {and} \bibinfo{person}{Amy
  Bruckman}.} \bibinfo{year}{2021}\natexlab{}.
\newblock \showarticletitle{Evaluating the {Effectiveness} of {Deplatforming}
  as a {Moderation} {Strategy} on {Twitter}}.
\newblock \bibinfo{journal}{\emph{Proceedings of the ACM on Human-Computer
  Interaction}} \bibinfo{volume}{5}, \bibinfo{number}{CSCW2}
  (\bibinfo{date}{Oct.} \bibinfo{year}{2021}), \bibinfo{pages}{1--30}.
\newblock
\showISSN{2573-0142}
\urldef\tempurl%
\url{https://doi.org/10.1145/3479525}
\showDOI{\tempurl}


\bibitem[Jhaver et~al\mbox{.}(2019c)]%
        {jhaver2019does}
\bibfield{author}{\bibinfo{person}{Shagun Jhaver}, \bibinfo{person}{Amy
  Bruckman}, {and} \bibinfo{person}{Eric Gilbert}.}
  \bibinfo{year}{2019}\natexlab{c}.
\newblock \showarticletitle{Does transparency in moderation really matter? User
  behavior after content removal explanations on reddit}.
\newblock \bibinfo{journal}{\emph{Proceedings of the ACM on Human-Computer
  Interaction}} \bibinfo{volume}{3}, \bibinfo{number}{CSCW}
  (\bibinfo{year}{2019}), \bibinfo{pages}{1--27}.
\newblock


\bibitem[Jhaver et~al\mbox{.}(2018)]%
        {jhaver_online_2018}
\bibfield{author}{\bibinfo{person}{Shagun Jhaver}, \bibinfo{person}{Sucheta
  Ghoshal}, \bibinfo{person}{Amy Bruckman}, {and} \bibinfo{person}{Eric
  Gilbert}.} \bibinfo{year}{2018}\natexlab{}.
\newblock \showarticletitle{Online {Harassment} and {Content} {Moderation}:
  {The} {Case} of {Blocklists}}.
\newblock \bibinfo{journal}{\emph{ACM Transactions on Computer-Human
  Interaction}} \bibinfo{volume}{25}, \bibinfo{number}{2}
  (\bibinfo{date}{April} \bibinfo{year}{2018}), \bibinfo{pages}{1--33}.
\newblock
\showISSN{1073-0516, 1557-7325}
\urldef\tempurl%
\url{https://doi.org/10.1145/3185593}
\showDOI{\tempurl}


\bibitem[Keegan and Fiesler(2017)]%
        {Brian_keegan_2017}
\bibfield{author}{\bibinfo{person}{Brian Keegan} {and} \bibinfo{person}{Casey
  Fiesler}.} \bibinfo{year}{2017}\natexlab{}.
\newblock \showarticletitle{The Evolution and Consequences of Peer Producing
  Wikipedia's Rules}. In \bibinfo{booktitle}{\emph{Proceedings of ICWSM}}.
\newblock
\urldef\tempurl%
\url{https://aaai.org/ocs/index.php/ICWSM/ICWSM17/paper/view/15698}
\showURL{%
\tempurl}


\bibitem[Keller et~al\mbox{.}(2020)]%
        {keller2020facts}
\bibfield{author}{\bibinfo{person}{Daphne Keller}, \bibinfo{person}{Paddy
  Leerssen}, {et~al\mbox{.}}} \bibinfo{year}{2020}\natexlab{}.
\newblock \showarticletitle{Facts and where to find them: Empirical research on
  internet platforms and content moderation}.
\newblock \bibinfo{journal}{\emph{Social media and democracy: The state of the
  field and prospects for reform}}  \bibinfo{volume}{220}
  (\bibinfo{year}{2020}), \bibinfo{pages}{224}.
\newblock


\bibitem[Kiesler et~al\mbox{.}(2011)]%
        {kiesler2011regulating}
\bibfield{author}{\bibinfo{person}{Sara Kiesler}, \bibinfo{person}{Robert~E.
  Kraut}, \bibinfo{person}{Paul Resnick}, \bibinfo{person}{Aniket Kittur},
  \bibinfo{person}{Moira Burke}, \bibinfo{person}{Yan Chen},
  \bibinfo{person}{Niki Kittur}, \bibinfo{person}{Joseph Konstan},
  \bibinfo{person}{Yuqing Ren}, {and} \bibinfo{person}{John Riedl}.}
  \bibinfo{year}{2011}\natexlab{}.
\newblock \bibinfo{booktitle}{\emph{Regulating Behavior in Online
  Communities}}.
\newblock \bibinfo{publisher}{The MIT Press}, \bibinfo{pages}{125--178}.
\newblock
\showISBNx{9780262016575}
\urldef\tempurl%
\url{http://www.jstor.org/stable/j.ctt5hhgvw.7}
\showURL{%
\tempurl}


\bibitem[Klonick(2017)]%
        {klonick2017new}
\bibfield{author}{\bibinfo{person}{Kate Klonick}.}
  \bibinfo{year}{2017}\natexlab{}.
\newblock \showarticletitle{The new governors: The people, rules, and processes
  governing online speech}.
\newblock \bibinfo{journal}{\emph{Harv. L. Rev.}}  \bibinfo{volume}{131}
  (\bibinfo{year}{2017}), \bibinfo{pages}{1598}.
\newblock


\bibitem[{Knight, Ben}(2018)]%
        {knight18}
\bibfield{author}{\bibinfo{person}{{Knight, Ben}}.}
  \bibinfo{year}{2018}\natexlab{}.
\newblock \bibinfo{title}{Germany implements new internet hate speech
  crackdown}.
\newblock
  \bibinfo{howpublished}{\url{https://www.dw.com/en/germany-implements-new-internet-hate-speech-crackdown/a-41991590}}.
\newblock


\bibitem[langdetect(2023)]%
        {langdetect}
langdetect \bibinfo{year}{2023}\natexlab{}.
\newblock
\newblock
\urldef\tempurl%
\url{https://github.com/Mimino666/langdetect}
\showURL{%
\tempurl}


\bibitem[{Le Pochat} et~al\mbox{.}(2019)]%
        {LePochat2019}
\bibfield{author}{\bibinfo{person}{Victor {Le Pochat}}, \bibinfo{person}{Tom
  {Van Goethem}}, \bibinfo{person}{Samaneh Tajalizadehkhoob},
  \bibinfo{person}{Maciej Korczy\'{n}ski}, {and} \bibinfo{person}{Wouter
  Joosen}.} \bibinfo{year}{2019}\natexlab{}.
\newblock \showarticletitle{Tranco: A Research-Oriented Top Sites Ranking
  Hardened Against Manipulation}. In \bibinfo{booktitle}{\emph{Proceedings of
  the 26th Annual Network and Distributed System Security Symposium}}
  \emph{(\bibinfo{series}{NDSS 2019})}.
\newblock
\urldef\tempurl%
\url{https://doi.org/10.14722/ndss.2019.23386}
\showDOI{\tempurl}


\bibitem[Loomba et~al\mbox{.}(2021)]%
        {loomba2021measuring}
\bibfield{author}{\bibinfo{person}{Sahil Loomba}, \bibinfo{person}{Alexandre de
  Figueiredo}, \bibinfo{person}{Simon~J Piatek}, \bibinfo{person}{Kristen de
  Graaf}, {and} \bibinfo{person}{Heidi~J Larson}.}
  \bibinfo{year}{2021}\natexlab{}.
\newblock \showarticletitle{Measuring the impact of COVID-19 vaccine
  misinformation on vaccination intent in the UK and USA}.
\newblock \bibinfo{journal}{\emph{Nature human behaviour}} \bibinfo{volume}{5},
  \bibinfo{number}{3} (\bibinfo{year}{2021}), \bibinfo{pages}{337--348}.
\newblock


\bibitem[{Mackintosh, Eliza}(2021)]%
        {mackintosh21}
\bibfield{author}{\bibinfo{person}{{Mackintosh, Eliza}}.}
  \bibinfo{year}{2021}\natexlab{}.
\newblock \bibinfo{title}{Facebook knew it was being used to incite violence in
  Ethiopia. It did little to stop the spread, documents show}.
\newblock
  \bibinfo{howpublished}{\url{https://www.cnn.com/2021/10/25/business/ethiopia-violence-facebook-papers-cmd-intl/index.html}}.
\newblock


\bibitem[Matias(2016)]%
        {matias_going_2016}
\bibfield{author}{\bibinfo{person}{J.~Nathan Matias}.}
  \bibinfo{year}{2016}\natexlab{}.
\newblock \showarticletitle{Going {Dark}: {Social} {Factors} in {Collective}
  {Action} {Against} {Platform} {Operators} in the {Reddit} {Blackout}}. In
  \bibinfo{booktitle}{\emph{Proceedings of the 2016 {CHI} {Conference} on
  {Human} {Factors} in {Computing} {Systems}}}. \bibinfo{publisher}{ACM},
  \bibinfo{address}{San Jose California USA}, \bibinfo{pages}{1138--1151}.
\newblock
\showISBNx{978-1-4503-3362-7}
\urldef\tempurl%
\url{https://doi.org/10.1145/2858036.2858391}
\showDOI{\tempurl}


\bibitem[Matias(2019)]%
        {matias2019preventing}
\bibfield{author}{\bibinfo{person}{J.~Nathan Matias}.}
  \bibinfo{year}{2019}\natexlab{}.
\newblock \showarticletitle{Preventing harassment and increasing group
  participation through social norms in 2,190 online science discussions}.
\newblock \bibinfo{journal}{\emph{Proceedings of the National Academy of
  Sciences}} \bibinfo{volume}{116}, \bibinfo{number}{20}
  (\bibinfo{year}{2019}), \bibinfo{pages}{9785--9789}.
\newblock


\bibitem[Matias et~al\mbox{.}(2022)]%
        {matias2022software}
\bibfield{author}{\bibinfo{person}{J~Nathan Matias}, \bibinfo{person}{Austin
  Hounsel}, {and} \bibinfo{person}{Nick Feamster}.}
  \bibinfo{year}{2022}\natexlab{}.
\newblock \showarticletitle{Software-supported audits of decision-making
  systems: Testing google and facebook's political advertising policies}.
\newblock \bibinfo{journal}{\emph{Proceedings of the ACM on Human-Computer
  Interaction}} \bibinfo{volume}{6}, \bibinfo{number}{CSCW1}
  (\bibinfo{year}{2022}), \bibinfo{pages}{1--19}.
\newblock


\bibitem[Myers~West(2018)]%
        {myers_west_censored_2018}
\bibfield{author}{\bibinfo{person}{Sarah Myers~West}.}
  \bibinfo{year}{2018}\natexlab{}.
\newblock \showarticletitle{Censored, suspended, shadowbanned: {User}
  interpretations of content moderation on social media platforms}.
\newblock \bibinfo{journal}{\emph{New Media \& Society}} \bibinfo{volume}{20},
  \bibinfo{number}{11} (\bibinfo{date}{Nov.} \bibinfo{year}{2018}),
  \bibinfo{pages}{4366--4383}.
\newblock
\showISSN{1461-4448, 1461-7315}
\urldef\tempurl%
\url{https://doi.org/10.1177/1461444818773059}
\showDOI{\tempurl}


\bibitem[Ozanne et~al\mbox{.}(2022)]%
        {ozanne2022shall}
\bibfield{author}{\bibinfo{person}{Marie Ozanne}, \bibinfo{person}{Aparajita
  Bhandari}, \bibinfo{person}{Natalya~N Bazarova}, {and}
  \bibinfo{person}{Dominic DiFranzo}.} \bibinfo{year}{2022}\natexlab{}.
\newblock \showarticletitle{Shall AI moderators be made visible? Perception of
  accountability and trust in moderation systems on social media platforms}.
\newblock \bibinfo{journal}{\emph{Big Data \& Society}} \bibinfo{volume}{9},
  \bibinfo{number}{2} (\bibinfo{year}{2022}),
  \bibinfo{pages}{20539517221115666}.
\newblock


\bibitem[Roberts(2014)]%
        {roberts2014behind}
\bibfield{author}{\bibinfo{person}{Sarah~T Roberts}.}
  \bibinfo{year}{2014}\natexlab{}.
\newblock \bibinfo{booktitle}{\emph{Behind the screen: The hidden digital labor
  of commercial content moderation}}.
\newblock \bibinfo{publisher}{University of Illinois at Urbana-Champaign}.
\newblock


\bibitem[Roberts(2016)]%
        {roberts_commercial_nodate}
\bibfield{author}{\bibinfo{person}{Sarah~T Roberts}.}
  \bibinfo{year}{2016}\natexlab{}.
\newblock \showarticletitle{Commercial content moderation: Digital laborers'
  dirty work}.
\newblock  (\bibinfo{year}{2016}).
\newblock


\bibitem[{Sabin, Sam}(2019)]%
        {sabin23}
\bibfield{author}{\bibinfo{person}{{Sabin, Sam}}.}
  \bibinfo{year}{2019}\natexlab{}.
\newblock \bibinfo{title}{On Policing Content, Social Media Companies Face a
  Trust Gap With Users}.
\newblock
  \bibinfo{howpublished}{\url{https://pro.morningconsult.com/articles/on-policing-content-social-media-companies-face-a-trust-gap-with-users}}.
\newblock


\bibitem[Salda{\~n}a(2021)]%
        {saldana2021coding}
\bibfield{author}{\bibinfo{person}{Johnny Salda{\~n}a}.}
  \bibinfo{year}{2021}\natexlab{}.
\newblock \bibinfo{booktitle}{\emph{The coding manual for qualitative
  researchers}}.
\newblock \bibinfo{publisher}{sage}.
\newblock


\bibitem[Sch{\"a}fer(2002)]%
        {schafer2002legal}
\bibfield{author}{\bibinfo{person}{Hans-Bernd Sch{\"a}fer}.}
  \bibinfo{year}{2002}\natexlab{}.
\newblock \showarticletitle{Legal rules and standards}.
\newblock In \bibinfo{booktitle}{\emph{The Encyclopedia of Public Choice}}.
  \bibinfo{publisher}{Springer}, \bibinfo{pages}{671--674}.
\newblock


\bibitem[{Scrapy | A Fast and Powerful Scraping and Web Crawling
  Framework}(2023)]%
        {scrapy}
{Scrapy | A Fast and Powerful Scraping and Web Crawling Framework}
  \bibinfo{year}{2023}\natexlab{}.
\newblock
\newblock
\urldef\tempurl%
\url{https://scrapy.org/}
\showURL{%
\tempurl}


\bibitem[Seering et~al\mbox{.}(2017)]%
        {seering_shaping_2017}
\bibfield{author}{\bibinfo{person}{Joseph Seering}, \bibinfo{person}{Robert
  Kraut}, {and} \bibinfo{person}{Laura Dabbish}.}
  \bibinfo{year}{2017}\natexlab{}.
\newblock \showarticletitle{Shaping pro and anti-social behavior on {Twitch}
  through moderation and example-setting}. In
  \bibinfo{booktitle}{\emph{Proceedings of CSCW}}. \bibinfo{publisher}{{ACM}},
  \bibinfo{address}{New York, NY, USA}.
\newblock


\bibitem[Singhal et~al\mbox{.}(2023)]%
        {singhal2022sok}
\bibfield{author}{\bibinfo{person}{Mohit Singhal}, \bibinfo{person}{Chen Ling},
  \bibinfo{person}{Nihal Kumarswamy}, \bibinfo{person}{Gianluca Stringhini},
  {and} \bibinfo{person}{Shirin Nilizadeh}.} \bibinfo{year}{2023}\natexlab{}.
\newblock \showarticletitle{SoK: Content moderation in social media, from
  guidelines to enforcement, and research to practice}. In
  \bibinfo{booktitle}{\emph{8th IEEE European Symposium on Security and Privacy
  (EuroSP 2023)}}.
\newblock


\bibitem[Software(2021)]%
        {maxqda}
\bibfield{author}{\bibinfo{person}{VERBI Software}.}
  \bibinfo{year}{2021}\natexlab{}.
\newblock \bibinfo{title}{{MAXQDA}}.
\newblock
\newblock
\urldef\tempurl%
\url{https://www.maxqda.com/}
\showURL{%
\tempurl}


\bibitem[Srinivasan et~al\mbox{.}(2019)]%
        {srinivasan+dnm+lee+tan:19}
\bibfield{author}{\bibinfo{person}{Kumar~Bhargav Srinivasan},
  \bibinfo{person}{Cristian Danescu-Niculescu-Mizil}, \bibinfo{person}{Lillian
  Lee}, {and} \bibinfo{person}{Chenhao Tan}.} \bibinfo{year}{2019}\natexlab{}.
\newblock \showarticletitle{Content removal as a moderation strategy:
  Compliance and other outcomes in the ChangeMyView community}. In
  \bibinfo{booktitle}{\emph{Proceedings of CSCW}}.
\newblock


\bibitem[Suzor(2019)]%
        {suzor2019lawless}
\bibfield{author}{\bibinfo{person}{Nicolas~P Suzor}.}
  \bibinfo{year}{2019}\natexlab{}.
\newblock \bibinfo{booktitle}{\emph{Lawless: The secret rules that govern our
  digital lives}}.
\newblock \bibinfo{publisher}{Cambridge University Press}.
\newblock


\bibitem[Taylor(2023)]%
        {tiktok_replace_twitter}
\bibfield{author}{\bibinfo{person}{Lorenz Taylor}.}
  \bibinfo{year}{2023}\natexlab{}.
\newblock \bibinfo{title}{{How Twitter lost its place as the global town
  square}}.
\newblock
\newblock
\urldef\tempurl%
\url{https://www.washingtonpost.com/technology/2023/07/07/twitter-dead-musk-tiktok-public-square/}
\showURL{%
\tempurl}


\bibitem[undetected-chromedriver(2023)]%
        {ud-chrome}
undetected-chromedriver \bibinfo{year}{2023}\natexlab{}.
\newblock
\newblock
\urldef\tempurl%
\url{https://github.com/ultrafunkamsterdam/undetected-chromedriver}
\showURL{%
\tempurl}


\bibitem[Vaccaro et~al\mbox{.}(2020)]%
        {vaccaro_at_2020}
\bibfield{author}{\bibinfo{person}{Kristen Vaccaro}, \bibinfo{person}{Christian
  Sandvig}, {and} \bibinfo{person}{Karrie Karahalios}.}
  \bibinfo{year}{2020}\natexlab{}.
\newblock \showarticletitle{"{At} the {End} of the {Day} {Facebook} {Does}
  {What} {ItWants}": {How} {Users} {Experience} {Contesting} {Algorithmic}
  {Content} {Moderation}}.
\newblock \bibinfo{journal}{\emph{Proceedings of the ACM on Human-Computer
  Interaction}} \bibinfo{volume}{4}, \bibinfo{number}{CSCW2}
  (\bibinfo{date}{Oct.} \bibinfo{year}{2020}), \bibinfo{pages}{1--22}.
\newblock
\showISSN{2573-0142}
\urldef\tempurl%
\url{https://doi.org/10.1145/3415238}
\showDOI{\tempurl}


\bibitem[{Williams, Jamie}(2018)]%
        {sandvig18}
\bibfield{author}{\bibinfo{person}{{Williams, Jamie}}.}
  \bibinfo{year}{2018}\natexlab{}.
\newblock \bibinfo{title}{D.C. Court: Accessing Public Information is Not a
  Computer Crime}.
\newblock
  \bibinfo{howpublished}{\url{https://www.eff.org/deeplinks/2018/04/dc-court-accessing-public-information-not-computer-crime}}.
\newblock


\bibitem[Wilson et~al\mbox{.}(2016)]%
        {wilson2016creation}
\bibfield{author}{\bibinfo{person}{Shomir Wilson}, \bibinfo{person}{Florian
  Schaub}, \bibinfo{person}{Aswarth~Abhilash Dara}, \bibinfo{person}{Frederick
  Liu}, \bibinfo{person}{Sushain Cherivirala}, \bibinfo{person}{Pedro~Giovanni
  Leon}, \bibinfo{person}{Mads~Schaarup Andersen}, \bibinfo{person}{Sebastian
  Zimmeck}, \bibinfo{person}{Kanthashree~Mysore Sathyendra},
  \bibinfo{person}{N~Cameron Russell}, {et~al\mbox{.}}}
  \bibinfo{year}{2016}\natexlab{}.
\newblock \showarticletitle{The creation and analysis of a website privacy
  policy corpus}. In \bibinfo{booktitle}{\emph{Proceedings of the 54th Annual
  Meeting of the Association for Computational Linguistics (Volume 1: Long
  Papers)}}. \bibinfo{pages}{1330--1340}.
\newblock


\bibitem[Zannettou(2021)]%
        {zannettou_i_2021}
\bibfield{author}{\bibinfo{person}{Savvas Zannettou}.}
  \bibinfo{year}{2021}\natexlab{}.
\newblock \showarticletitle{"{I} {Won} the {Election}!": {An} {Empirical}
  {Analysis} of {Soft} {Moderation} {Interventions} on {Twitter}}.
\newblock \bibinfo{journal}{\emph{arXiv:2101.07183 [cs]}}
  (\bibinfo{date}{April} \bibinfo{year}{2021}).
\newblock
\urldef\tempurl%
\url{http://arxiv.org/abs/2101.07183}
\showURL{%
\tempurl}
\newblock
\shownote{arXiv: 2101.07183}.


\end{thebibliography}
